\documentclass[11pt,a4paper]{article}
\pdfoutput = 1
\usepackage{jcappub}
\usepackage[utf8]{inputenc}
\usepackage[T1]{fontenc}
\usepackage{amsmath,amssymb,bm}
\usepackage{graphicx}
\usepackage{hyperref}
\usepackage{natbib}
\usepackage{caption}
\usepackage{subcaption}
\usepackage{xcolor}

\subheader{
\footnotesize
\vspace*{-3.7em}
\begin{flushright}
RBI-ThPhys-2023-22
\end{flushright}
}

\renewcommand{\vec}[1]{\bm{#1}}
\newcommand*{\pp}  {\parallel}

\newcommand*{\df}  {\delta}

\newcommand*{\non}  {\nonumber}
\newcommand*{\lb}  {\left(}
\newcommand*{\rb}  {\right)}
\newcommand*{\ls}  {\left[}
\newcommand*{\rs}  {\right]}
\newcommand*{\la}  {\left\langle}
\newcommand*{\ra}  {\right\rangle}

\newcommand{\vx} {\vec{x}}
\newcommand{\vk} {\vec{k}}

\newcommand{\ba}{\[\begin{aligned}}
\newcommand{\ea}{\end{aligned}\]}
\newcommand{\eq}[1]{\begin{align}#1\end{align}}
\newcommand{\eeq}[1]{\begin{equation}#1\end{equation}}
\newcommand{\ZG}[1]{{#1}}

\author[a,b]{Zucheng Gao}
\author[c,b,d]{Zvonimir Vlah}
\author[a,b,d]{Anthony Challinor}

\affiliation[a]{Institute of Astronomy, Madingley Road, Cambridge, CB3 0HA, UK,}
\affiliation[b]{Kavli Institute for Cosmology Cambridge, Madingley Road, Cambridge, CB3 0HA, UK,}
\affiliation[c]{Division of Theoretical Physics, 
Ru\dj er Bo\v{s}kovi\'{c} Institute, Zagreb HR-10000, Croatia,}
\affiliation[d]{DAMTP, Centre for Mathematical Sciences, Wilberforce Road, Cambridge CB3 0WA, UK.}

\emailAdd{zg285@ast.cam.ac.uk}
\emailAdd{zvlah@irb.hr}
\emailAdd{a.d.challinor@ast.cam.ac.uk}

\title{Flat-sky angular power spectra revisited}


\abstract{
We revisit the flat-sky approximation for evaluating the angular power spectra of projected random fields by retaining information about the correlations along the line of sight. For the case of projections with broad, overlapping radial window functions, these line-of-sight correlations are suppressed and are ignored in the commonly adopted Limber approximation. However, retaining the correlations is important for narrow window functions or unequal-time spectra but introduces significant computational difficulties due to the highly oscillatory nature of the integrands involved. We deal with the integral over line-of-sight wave-modes in the flat-sky approximation analytically, using the FFTlog expansion of the 3D power spectrum.
This results in an efficient computational method, 
which is a substantial improvement compared to any full-sky approaches. We apply our results to galaxy clustering (with and without redshift-space distortions), CMB lensing and galaxy lensing observables in a flat $\Lambda\text{CDM}$ universe. In the case of galaxy clustering, we find excellent agreement with the full-sky results on large (percent-level agreement) and intermediate or small (subpercent agreement) scales, dramatically out-performing the Limber approximation for both wide and narrow window functions, and in equal- and unequal-time cases. In the cases of lensing, we show on the full-sky that the angular power spectrum of the lensing convergence can be very well approximated by projecting the 3D Laplacian (rather than the correct angular Laplacian) of the gravitational potential, even on large scales. Combining this approximation with our flat-sky techniques provides an efficient and accurate evaluation of the CMB lensing angular power spectrum on all scales.
We further analyse the clustering and lensing angular power spectra by isolating the projection effects due to the observable- and survey-specific window functions, separating them from the effects due to integration along the line of sight and unequal-time mixing in the 3D power spectrum. All of the angular power spectrum results presented in this paper are obtained using a Python code implementation, which we make publicly available.
}

\begin{document}

\maketitle
\flushbottom

\section{Introduction}
\label{sec:intro}

Next-generation galaxy surveys, such as Euclid~\cite{Amendola:2016}, DESI~\cite{Aghamousa:2016}, Rubin~\cite{Abate:2012}, Roman~\cite{Spergel:2015}, SPHEREx~\cite{Dore2014, Dore2018} and others, aim to address a range of cosmological questions, from uncovering the nature of dark energy and tests of general relativity on large scales~\cite{Weinberg:2013,Mortonson:2013,Munshi:2015}, to constraining the properties of the initial conditions of the universe by measuring signals of primordial non-Gaussianity \cite{Linde1997, Maldacena2003, Dalal2008, Afshordi2008, Matarrese2008}. To succeed in these tasks, reliable, accurate and efficient measurements of the galaxy-overdensity statistics are paramount, amongst which the two-point functions (e.g., angular spectra and 3D power spectra) take a leading role. 

The angular power spectrum has been an observable of choice in many surveys, especially for the study of weak gravitational lensing and the cosmic microwave background (CMB), where the extensive comparisons of theoretical predictions and observations are used with the goal of constraining cosmological parameters. Even for galaxy clustering surveys, where the 3D power spectrum is the most commonly used statistic, the angular power spectrum has certain advantages, e.g., it is defined in terms of variables -- redshifts and angular coordinates -- which are cosmology independent. Moreover, on large angular scales, the standard 3D power spectrum analysis starts to exhibit effects related to the fixed line of sight. On the other hand, angular power spectra are naturally defined on the full sky. However, the evaluation of full-sky angular power spectra is computationally demanding (see, e.g., \cite{Pratten:2013, Reimberg:2015, Gebhardt:2021, Raccanelli_I:2023, Gao:2023} for discussion), as it involves integration over two spherical Bessel functions, making the integrands highly oscillatory. Furthermore, for spectroscopic surveys many angular power spectra are required between observables projected in narrow radial window functions to avoid loss of information.
There have been several attempts to speed up the full-sky evaluation based on the FFTLog algorithm~\cite{Hamilton2000, Assassi2017, Schoneberg:2018, Fang:2020}: by expanding the 3D power spectrum in (complex) powers of the wavenumber $k$, the integration over spherical Bessel functions can be performed analytically in terms of special functions. This provides significant improvements relative to direct computations, but the evaluation still poses substantial computational challenges and thus motivates the search for alternative approaches. One commonly used approach relies on a set of approximations resulting in the `Limber approximation' solution \cite{Limber:1954, Peebles:1973}, generically accurate on small scales (large multipoles $\ell$) and appropriate only for broad and overlapping radial window functions. However, relying on the Limber approximation for galaxy clustering in future surveys could lead to a biased cosmological analysis (e.g., \cite{Fang:2020}), particularly if the focus is on large or intermediate scales to avoid other modelling challenges (e.g., scale-dependent bias, non-linear clustering and baryonic effects). For these reasons, a middle-ground between the full-sky treatment and Limber-like approximations may be useful. Furthermore, next-generation surveys will exhibit improved photometric-redshift accuracy in clustering measurements with narrower radial window functions, allowing efficient cross-correlation of galaxy fields. This is precisely the regime where the Limber approximation is known not to be valid and thus cannot be relied upon.

An intermediate regime between the full-sky results and Limber approximations has received much less attention in the literature, with the existing works varying in the degree of approximations and corrections they consider \cite{Datta:2006, Afshordi2008, Reimberg:2015, White:2017, Castorina:2017, Castorina:2018, Jalilvand:2019}.
The difference between Limber and these flat-sky results lies in the treatment of the correlations along the line-of-sight, which Limber neglects. Recently, ref.~\cite{Matthewson:2020} explored the accuracy of the flat-sky approximation by comparing it to the full-sky results in various setups, finding that it can reach subpercent agreement in galaxy number counts and galaxy lensing scenarios. We highlight that the precise form of these various flat-sky approximations, presented in the literature, can differ in certain details that can be traced to the different choices in the geometric and other approximations. Most of the approaches choose the 3D two-point correlation function as the starting point. Moreover, these differences tend to become starker in the unequal-time case (when cross-correlating galaxy sources from different redshift bins), and expressions often get increasingly more complex compared to the equal-time case.

\begin{table*}
\centering
\begin{tabular}{ll}
\hline
\hline
Symbol & Definition \\
\hline
$\df^{\rm K}_{ij}$ & Kronecker symbol \\[3pt]
$\df^{n\rm D} (\vec{x})$ & Dirac delta function in $n$ dimensions\\
$ W(\chi) $ & Radial window function; related to the specific observable and survey\\
\hline
$\df(\vec{x})$ & 3D over-density field of matter or biased tracer \\
$\hat \df(\theta, \phi)$ & 2D projected field in angular coordinates on the sky  \\
\hline
$P(k;\, \chi, \chi')$ & Unequal-time power spectrum of the 3D density field \\
$\mathbb{C}_{\ell}  (\chi,\chi')$ & Unequal-time angular power spectrum (in the narrow window function limit) \\
$C_{\ell}$ & Projected angular power spectrum (with finite width window functions) \\
\hline
\end{tabular}
\caption{Notation used for the most important quantities throughout this paper.}
\label{tab:notation}
\end{table*}

In this work, we revisit the flat-sky approximation enhancing the treatment in several ways:
(i) we consider the case of unequal-time projections and discuss how to define a single effective radius for connecting the scale of transverse spatial fluctations to angular fluctuations;
(ii) we optimize the flat-sky calculation using the FFTLog algorithm, which can greatly speed up the calculation; and (iii) we compare our results to the full-sky results in the case of galaxy clustering and CMB lensing, providing a detailed analysis of where the flat-sky approximation deviates from the full-sky. We investigate the dependence on 3D scales $k$ of the angular power spectrum for each of these observables, analysing the importance of contributions from different integration regions before and after projection over the radial window functions. 

The paper is organized as follows. In section~\ref{sec:ang_ps}, after reviewing the full-sky angular power spectrum and Limber approximation, we derive the corresponding result within the flat-sky approximation retaining integration over the line-of-sight wave-modes. We present two independent derivations of these flat-sky results, starting in Fourier space or real space, respectively. The latter provides and motivates the commonly used geometric recalibration of scales in the flat-sky approximation, which we adopt for our numerical results.
In the same section, we describe our implementation of the FFTLog algorithm for a fast and optimized flat-sky calculation. In section~\ref{Sec:RD}, we apply our results to galaxy clustering (including also redshift-space distortions) and CMB lensing spectra. Section~\ref{Sec:discussion} provides a detailed discussion and comparison of the full- and flat-sky results. We also perform a simple asymptotic analysis of our flat-sky angular power spectrum clarifying its functional dependencies. We conclude in section~\ref{Sec:conclusion}. In appendix~\ref{app:full_sky_Cell} we give a short review of the full-sky angular power spectrum results, while appendix~\ref{app:kapparadialderivs} provides further discussion of approximations we adopt for lensing power spectra.

Throughout this paper, we work with the \textit{Planck} best-fit spatially flat $\Lambda\text{CDM}$ cosmology \cite{Aghanim:2018}, with CDM density $\Omega_c h^2 = 0.11933$, baryon density $\Omega_b h^2 = 0.02242$, Hubble constant $H_0 = 100 h \, \text{km\,s}^{-1}\,\text{Mpc}^{-1}$ with $h=0.6766$, scalar spectral index $n_s=0.9665$, and fluctuation amplitude $\sigma_8 = 0.8103$. In table~\ref{tab:notation} we summarise our notation for the key quantities that feature throughout the paper. All results presented here are derived with a Python code that we have developed and that we make publicly available on the GitHub repository.\footnote{https://github.com/GZCPhysics/BeyondLimber.git} It is built using Python3~\cite{Rossum:2009}, while all the basic cosmological functions are directly imported from the Boltzmann codes CAMB~\cite{Lewis:1999} (for calculation of the matter power spectrum and comparisons with the full-sky angular power spectrum of the CMB lensing potential)
and CLASS~\cite{Lesgourgues:2011} (for calculation of the linear growth factor and logarithmic growth rate).

\section{Angular power spectra of projected 3D fields}
\label{sec:ang_ps}

In this section, we study the angular power spectrum for a general observable obtained by projecting a 3D random field. We start with a short review of the full-sky results and the commonly used Limber approximation, also stating the assumptions that the latter is based on. In the rest of the section, we then derive the expression for the flat-sky approximation for the angular power spectrum that, crucially, relies on incorporating the wave-modes along the line of sight. We also discuss the origin of the geometrical recalibration of multipoles $\ell$, which is often made in the Limber approximation. Finally, we present our efficient numerical algorithm, based on FFTlog, for evaluation of angular power spectra in the flat-sky approximation.

\subsection{Full-sky angular power spectrum}
\label{subsec:full-sky}

We consider the projection $\hat\delta(\theta, \phi)$ of some 3D field $\delta$, where $\theta$ and $\phi$ are spherical coordinates. It is convenient to use the spherical-harmonic expansion of $\hat\delta$,
\eeq{
 \hat \df (\theta, \phi) = \sum_{\ell = 0}^\infty \sum_{m=-\ell}^\ell \hat\delta_{\ell,m} Y_{\ell}^m(\theta,\phi)\, ,
}
where the coefficients $\hat\delta_{\ell,m}$ are easily obtained from the orthogonality of the spherical harmonics: $\hat\delta_{\ell,m} = \int d\Omega\; Y^{m\ast}_{\ell}(\theta,\phi) \hat \df (\theta, \phi)$. We consider the projected field $\hat\delta(\theta, \phi)$ in terms of the 3D field $\df (\chi, \theta, \phi; \eta)$ in real space,
\eeq{
\label{Eq:Projection}
\hat \df (\theta, \phi) = \int d\chi\; W(\chi) \df (\chi, \theta, \phi;\, \eta_0-\chi) 
 = \int d\chi\; W(\chi) \int \frac{d^3 \boldsymbol{k}}{(2\pi)^3}\; e^{i\vec{k} \cdot \vec{x}}\df (\vec{k};\, \eta_0-\chi)\, ,
}
where $W(\chi)$ is a radial window function describing the observing characteristics of a certain tracer, as well as the specifics of the survey geometry, and $\eta_0-\chi$ is the conformal look-back time. The quantity $\df (\vec{k})$ is the Fourier transform\footnote{We use the following Fourier transform convention: $f(\vec{k}) \equiv \int d^3 {\vec{x}} \, f(\vec{x}) e^{-i\vec{k}\cdot\vec{x}}$.} of the 3D over-density field, and is related to the 3D power spectrum as
\eeq{
\label{Eq:3DPower_def}
\la \df(\vec{k};\, \eta_0-\chi) \df (\vec{k}';\, \eta_0-\chi')\ra = (2\pi)^3\df^{3\rm D} (\vec{k} + \vec{k}') P(k;\, \chi, \chi')\, .   
} 
For compactness, we use $\chi$ to label the look-back time $\eta_0-\chi$ in the power spectrum. To maintain generality, we are interested in the angular cross-power spectrum of projected fields $\hat{\delta}(\theta,\phi)$ and $\hat{\delta}'(\theta,\phi)$, where $\hat{\delta}'$ is constructed with a window function $W'(\chi)$; this is given by
\eq{
\la \hat{\delta}_{\ell,m} (\hat{\delta}_{\ell',m'}^\prime)^\ast \ra &= \int d\Omega d\Omega'\; Y^{m\ast}_{\ell}(\theta,\phi)
Y^{m'}_{\ell'}(\theta',\phi') \la \hat\df (\theta,\phi) \hat\delta^{\prime}(\theta',\phi') \ra \non \\
& = 4\pi  \df^{\rm K}_{mm'} \df^{\rm K}_{\ell \ell'} \int d\chi d\chi'\; W(\chi)W'(\chi')\; 
\int \frac{k^2 dk}{2\pi^2}\; j_{\ell}(k\chi)j_{\ell}(k\chi') P(k;\, \chi, \chi')\, ,  
\label{Eq:fullpower}
}
where $\df^{\rm K}$ denotes the Kronecker delta symbol. Note how the projected fields are statistically isotropic, with no correlations between different multipoles, a property inherited from the statistical homogeneity and isotropy of the 3D field being projected.
As usual, we define the angular power spectrum $C_{\ell}$ as
\eeq{
\la \hat{\delta}_{\ell,m} (\hat{\delta}_{\ell',m'}^\prime)^\ast \ra 
= \df^{\rm K}_{mm'} \df^{\rm K}_{\ell \ell'} C_{\ell}\, ,
}
and we have the relation between the angular power spectrum and the unequal-time 3D power spectrum
\eeq{
\label{Eq:full_sky}
C_{\ell} 
= \int d \chi d \chi'\, W(\chi) W'(\chi') \,\mathbb{C}_{\ell} (\chi,\chi')\, ,
}
where we have introduced the \textit{unequal-time angular power spectrum} $\mathbb{C}_{\ell} (\chi,\chi')$, 
given by
\eeq{
\label{Eq:curly_full}
\mathbb{C}_{\ell}  (\chi,\chi') \equiv 4\pi \int \frac{k^2 dk}{2\pi^2}\; P(k;\, \chi,\chi') \, j_\ell(k \chi) j_\ell(k \chi')\, .
}
The unequal-time angular power spectrum $\mathbb{C}_{\ell} $ can also be viewed as the usual projected angular power spectrum in the limit of narrow window functions, $W(\chi) \to \df^{\rm D}(\chi - \chi_*)$. Equation~\eqref{Eq:full_sky} cleanly separates the survey-specific radial selection, as encoded in the window functions, and the cosmology dependence, which is all contained in $\mathbb{C}_{\ell}$ through its dependence on the unequal-time 3D power spectrum and the mixing and projecting of 3D wave-modes and radial distances.
We will focus on these purely geometric and survey-independent projection effects in the next sections when we develop the flat-sky approximation. 

In the next sections, we will consider the unequal-time and projected angular power spectra in flat-sky and Limber approximations. In order to distinguish between the different quantities and their approximations, we introduce the label to the above spectra so that, in the full-sky case, we have $\mathbb{C}^{\rm full}_{\ell}(\chi,\chi')$ and $C^{\rm full}_{\ell}$ correspondingly for the unequal-time and projected angular power spectra.

\subsection{Limber approximation}
\label{subsec:limber}

In studying the statistical properties of extragalactic nebulae \cite{Limber:1954}, Limber first introduced 
several approximations to simplify the evaluation of the two-point angular correlation function 
(the inverse Legendre transform of the above angular power spectrum). Following this result, refs.~\cite{Peebles:1973, Kaiser:1992, Kaiser:1998} derived the Limber approximation directly in Fourier space. Due to its simplicity, involving a single integral for the Fourier-space version, the Limber approximation has remained one of the most commonly used means to evaluate the angular power spectrum. Its validity has been investigated in~\cite{Simon:2006,LoVerde2008,Lemos2017}.

There are two basic assumptions underlying the Limber approximation in Fourier space: (i) the sky is considered as flat; 
and (ii) the radial window functions are so broad compared to scales of interest for the inhomogeneities that modes with $k_{\parallel}\neq 0$ 
are suppressed to negligible levels by the radial integration. The second assumption is the one that
we will subsequently relax by including the physical effects of the modes along the line of sight.
We briefly re-derive the Limber approximation here, from these assumptions. Let $\vec{\theta}$ be angular position in the plane of the sky around some line of sight $\hat{\vec{n}}$, and $\hat{\delta}(\vec{\theta})$ the projected field at $\vec{\theta}$. Starting from eq.~(\ref{Eq:Projection}), the projected two-point correlation function in real space becomes
\eq{
\la \hat{\delta}(\vec{\theta}) \hat{\delta}'(\vec{\theta}') \ra 
& = \int d\chi d\chi'\; W(\chi)W'(\chi')\; \int \frac{d^3 \vec{k}}{(2\pi)^3}\; 
e^{i\vec{k} \cdot (\vec{x}-\vec{x}')} P(k;\, \chi, \chi')\, ,
}
where the 3D position $\vec{x} = \chi(\hat{\vec{n}}+\vec{\theta})$ and similarly for $\vec{x}'$.
Using the second assumption, i.e., neglecting the $k_{\parallel}$ contributions in the 3D
power spectrum $P(k;\, \chi, \chi') \approx P(|\vec{k}_\perp| ;\, \chi, \chi')$, we effectively  
constrain $\chi = \chi'$, and we can write 
\eeq{
\label{Eq:LimberCorrelation}
\la \hat{\delta}(\vec{\theta}) \hat{\delta}'(\vec{\theta}+\Delta\vec{\theta}) \ra 
= \int d\chi \; W(\chi)W'(\chi)\; \int \frac{d^2 \vec{k}_{\perp}}{(2\pi)^2}\; 
e^{i\chi \vec{k}_{\perp} \cdot \Delta \vec{\theta}} P(|\vec{k}_{\perp}|;\, \chi)\, .
}
The angular power spectrum in the Limber approximation is then easily obtained by performing the 2D Fourier transform:
\eq{
\label{Eq:LimberAngular}
C^{\rm Limber}_{\ell} 
= \int d^2\Delta \vec{\theta}\;  e^{-i\vec{\ell} \cdot \Delta \vec{\theta}} \la \hat{\delta}(\vec{\theta}) \hat{\delta}'(\vec{\theta}+\Delta\vec{\theta}) \ra 
= \int \frac{d\chi}{\chi^2} \; W(\chi)W'(\chi) P(\ell/\chi; \chi)\, .
}

We shall return to the Limber approximation in section~\ref{Sec:RD}, where we further discuss its validity in relation to exact, full-sky results and our new flat-sky approximation.

\subsection{Flat-sky angular power spectrum}
\label{subsec:OurApprox}

We now revisit the flat-sky angular power spectrum, but retaining the integration over modes along the line of sight.
By Fourier transforming the flat-sky projected field $\hat{\delta}(\vec{\theta})$, we obtain
\eq{
\hat \df (\vec{\ell}) & = \int d^2 \vec{\theta} \; e^{-i\vec{\ell} \cdot \vec{\theta}} \int d\chi\; W(\chi) \int \frac{d^3 \vec{k}}{(2\pi)^3}\; e^{i\chi\vec{k}_{\perp} \cdot \vec{\theta}} e^{ik_{\parallel}\chi}\df (\vec{k};\, \chi) \non \\
& = \int \frac{d\chi}{\chi^2}\; W(\chi) \int \frac{d k_{\parallel}}{2\pi}\; e^{ik_{\parallel}\chi}\df (k_{\parallel}\hat{\vec{n}}+\vec{\ell}/\chi;\, \chi)\, . 
\label{Eq:2DFourier}
}
Taking the two-point correlation of this quantity gives us 
\eq{
\label{Eq:chi1_chi2_power}
\la \hat \df (\vec{\ell}) \hat \df'(\vec{\ell}') \ra
&= \int \frac{d\chi}{\chi^2} \frac{d\chi'}{\chi'^2} ~ W(\chi)W'(\chi') \int \frac{d k_{\parallel}}{2\pi}\frac{d k'_{\parallel}}{2\pi}e^{ik_{\parallel}\chi}\, e^{ik'_{\parallel}\chi'}(2\pi)^3\delta^{\rm 3D}(\vec{k}+\vec{k}')P(k;\, \chi, \chi') \non \\
&= \int \frac{d\chi}{\chi^2} \frac{d\chi'}{\chi'^2} ~ W(\chi)W'(\chi') (2\pi)^2 
\df^{\rm 2D} \lb \vec{\ell}/\chi + \vec{\ell}'/\chi' \rb \non \\
&\hspace{2cm} \times  \int \frac{d k_{\parallel}}{2\pi} ~e^{i k_\parallel (\chi - \chi')} P \bigg( \sqrt{k_{\parallel}^2 +\frac{\ell \ell'}{\chi\chi'}};\, \chi, \chi' \bigg)\, ,
}
where, in the first line, $\vec{k} = k_{\parallel}\hat{\vec{n}}+\vec{\ell}/\chi$ and $\vec{k}' = k_{\parallel}\hat{\vec{n}}+\vec{\ell}'/\chi'$. 
Note that in the 3D power spectrum, we have set $k_\perp = \sqrt{\ell \ell'}/\sqrt{\chi \chi'}$ using the 2D Dirac delta function. This is convenient since it preserves the symmetry of the integral on the right under simultaneous exchange of $\vec{l}$ and $\vec{l}'$ and $W(\chi)$ and $W'(\chi')$, even without reference to the delta function, which will be important later when we approximate the delta function further.
Finally, changing the variables as $\chi = \bar{\chi} + \delta\chi/2$ and $\chi' = \bar{\chi} - \delta\chi/2$, we obtain
\eq{
\la \hat \df (\vec{\ell}) \hat \df'(\vec{\ell}') \ra
&= (2\pi)^2 \int \frac{d\bar{\chi}\, d \delta \chi}{ \left[\bar{\chi}^2 - (\delta \chi)^2/4 \right]^2 } ~ W\lb \bar{\chi} + \tfrac{1}{2} \delta \chi \rb W'\lb \bar{\chi} - \tfrac{1}{2} \delta \chi \rb
\df^{\rm 2D} \lb \vec{\ell}/\chi + \vec{\ell}'/\chi' \rb \non \\
& \hspace{1cm}\times \int \frac{d k_{\parallel}}{2\pi} ~ e^{ i k_{\parallel} \delta \chi}
P \bigg( \sqrt{k_{\parallel}^2 +\frac{\ell \ell'}{\chi\chi'}};\, \chi, \chi' \bigg)  \,.
\label{eq:flatcorrexpand}
}

Notice that in eq.~\eqref{eq:flatcorrexpand} the 2D Dirac delta function depends on the $\chi$ and $\chi'$ variables and thus cannot just be taken out of the radial integrals. The interpretation of this observation is that the unequal-time two-point correlation function, in the flat-sky approximation, is not diagonal in Fourier space, and there are also non-vanishing contributions when $\vec{\ell} + \vec{\ell}' \neq 0$. However, we saw that this does not happen in the full-sky case in eq.~\eqref{Eq:fullpower},
and is thus an artefact of performing the flat-sky projections at two different $\chi$s (unequal-time). We can understand this behaviour as follows. At radial distance $\chi$, translating $\vec{\theta}$ by $\vec{a}$ in the flat-sky approximation produces a transverse 3D displacement of $\chi \vec{a}$. The unequal-time two-point correlation function in real space will therefore not be a function of $|\vec{\theta} -\vec{\theta}'|$, but rather $|\chi\vec{\theta} - \chi'\vec{\theta}'|$ since this quantity is invariant under 3D transverse translations. The 2D delta function $\df^{\rm 2D} \lb \vec{\ell}/\chi + \vec{\ell}'/\chi' \rb$ in eq.~\eqref{eq:flatcorrexpand} is required to generate this dependence for the real-space angular correlation function.

Nonetheless, we are most interested in scenarios when the two variables $\chi$ and $\chi'$
and close to the mean, i.e., $\chi \approx \chi' \approx \bar \chi \gg \delta \chi$. As argued in \cite{Gao:2023, Raccanelli_II:2023}, we can formally expand the delta function around $\df^{\rm 2D} \lb \vec{\ell} + \vec{\ell}' \rb$ as
\eeq{
\df^{\rm 2D} \lb \vec{\ell}/{\chi} + \vec{\ell}'/{\chi'} \rb 
= \bar{\chi}^2 \lb 1 - \delta^2 \rb^2 ~ \df^{\rm 2D} \big( \vec{\ell} + \vec{\ell}' + \delta \vec{\Delta}  \big) = 
\bar{\chi}^2 \lb 1 - \delta^2 \rb^2 ~ e^{\delta \vec{\Delta} \cdot \partial_{\vec{\ell}'}} \df^{\rm 2D} 
\big( \vec{\ell} + \vec{\ell}'\big) \, ,
\label{eq:2D_delta}
}
where $\delta = \delta \chi/(2\bar{\chi})$ and $\vec{\Delta} = \vec{\ell}' - \vec{\ell}$. 
The two-point angular correlation function can thus be expanded around its 
diagonal component as
\eq{
\la \hat \df(\vec{\ell}) \hat \df'(\vec{\ell}') \ra
= (2\pi)^2
\df^{\rm 2D} \big( \vec{\ell} + \vec{\ell}' \big) \sum_{n=0}^\infty 
\frac{ ( \overset{\leftarrow}{\partial}_{\vec{\ell}'} \cdot \vec{\Delta})^n }{2^n n!} C^{(n)}(\sqrt{\ell\ell'}) \, ,
\label{eq:flatseries}
}
where the components of the projected angular spectra are given by
\eeq{
\label{Eq:Cln}
C^{(n)}(\ell) = \int d\bar{\chi}\, d \delta \chi \lb \frac{\delta \chi}{\bar{\chi}} \rb^n
W\lb \bar{\chi} + \tfrac{1}{2} \delta \chi \rb W'\lb \bar{\chi} - \tfrac{1}{2} \delta \chi \rb
\mathbb{C}^{\text{flat}}(\ell,\bar{\chi}, \delta \chi) \, , 
}
and we have introduced a flat-sky version of the unequal-time angular power spectrum
\eeq{
\label{Eq:curly_flat}
\mathbb{C}^{\text{flat}}(\ell,\bar{\chi}, \delta \chi) =  \frac{1}{\bar{\chi}^2}
\int \frac{d k_{\parallel}}{2\pi} ~ e^{ i k_{\parallel} \delta \chi}
P \left( \sqrt{k_{\parallel}^2 +\frac{\ell^2}{\bar{\chi}^2(1-\delta^2)}};\, \bar{\chi}, \delta \chi \right) \, . 
}
This can be compared to its full-sky counterpart given in eq.~\eqref{Eq:curly_full}. Recently, ref.~\cite{Gao:2023} showed that this flat-sky result can be obtained as a formal asymptotic limit of the full-sky case when $\ell$ is large and $|\delta \chi/\bar{\chi}|\propto 1/\ell$. In section~\ref{Sec:RD}, we will compare the two and show their close agreement in various cosmological scenarios. 

Regarding the higher components of the projected angular spectrum $C^{(n)}(\ell)$,
we can see that they are suppressed by the powers of $\delta \chi/\bar{\chi}$, which 
for suitable and compact choices of window functions $W(\chi)$ is an excellent expansion. 
In the rest of the paper, we thus consider only the diagonal piece $C^{\rm flat}(\ell) \equiv C^{(0)}(\ell)$
as our flat-sky approximation of the projected angular power spectrum. Note that for this term, we may replace the argument $\sqrt{\ell\ell'}$ of $C^{(0)}$ in eq.~\eqref{eq:flatseries} with $\ell$ since it is multiplied by the (undifferentiated) delta function.
One might be 
tempted to consider the higher components $C^{(n)}(\ell)$ as corrections to the $C^{\rm flat}(\ell)$
with the goal of providing better agreement with the full-sky result $C^{\rm full}_\ell$.
This is, however, not the correct interpretation. Indeed, as discussed earlier, from the set-up of our flat-sky 
result we have sacrificed the isotropy property (manifested as the translational invariance 
in the 2D plane) in the unequal-time case. The correct interpretation of the off-diagonal terms 
is thus as the estimate of the error on the validity of our flat-sky $C^{\rm flat}(\ell)$ result 
as an asymptotic approximation of the full-sky result $C^{\rm full}_\ell$, where the statistical 
isotropy is exactly realised.
 
At this point it is instructive to revisit the Limber approximation. As we have mentioned
in the previous subsection, in addition to the flat-sky approximation, the Limber approximation also neglects the dependence 
of the 3D power spectrum on the wave-modes along the line of sight.
Adopting this additional approximation, our result in equations~\eqref{Eq:Cln} and~\eqref{Eq:curly_flat} immediately reduces to
\begin{equation}
C^{(n)}(\ell) \rightarrow \delta^{\text{K}}_{n0} \int \frac{d\bar{\chi}}{\bar{\chi}^2} \; W(\bar{\chi})W'(\bar{\chi}) P(\ell/\bar{\chi}; \bar{\chi})\, ,
\end{equation}
in agreement with the Limber approximation (eq.~\ref{Eq:LimberAngular}).

We emphasise that our flat-sky results and the Limber approximation, therefore, differ only in
that we keep the non-zero $k_{\parallel}$ information, i.e., we retain the correlations 
along the line-of-sight. As we shall see, keeping these wave-modes is crucial
to capture features observed in the full-sky projected angular power spectrum,
especially for projected clustering at unequal times (see \cite{Raccanelli_II:2023} for the related discussion).

Before moving on, let us consider in more detail under which physical conditions we can expect the
Limber approximation to be applicable, i.e., when the wave-modes along the line of sight can be neglected (see also \cite{Simon:2006}). 
We start from the expression for $C^{\rm flat}$(i.e., $C^{(0)}$) given in eq.~\eqref{Eq:Cln}.
Neglecting the residual dependence on $\delta\chi$ in $\mathbb C^{\rm flat}$ [all except the term $\exp( i k_{\parallel} \delta \chi)$], 
we see that the wavenumber dependence of the 3D power spectrum
takes the form of $k^2 = k_\pp^2 + (\ell/\bar\chi)^2$. For the Limber assumption that $P(k)\approx P(\ell/\bar\chi)$ to hold, 
the effective $k_\pp$ domain has to be suppressed so that $\ell/\bar\chi\gg k_\pp$. This suppression can arise from the integral over $\delta \chi$ in the case of broad and overlapping radial window functions. If this integral is performed first, neglecting the sub-leading $\delta \chi$ contributions in the 3D power spectrum, it reduces to taking the Fourier transform of the $\delta \chi$-dependent part of the product of the two window functions.\footnote{For concreteness, let us consider 
a simple example of these $\delta \chi$ dependencies for equal Gaussian window functions, centred on $\chi_\ast$ and of width $\sigma$. We have
\eeq{
 W\lb \chi + \tfrac{1}{2} \delta \chi \rb W\lb \chi - \tfrac{1}{2} \delta \chi \rb 
= \frac{1}{2\pi \sigma^2} e^{- (\chi - \chi_*)^2/\sigma^2} e^{- \delta \chi^2/(4 \sigma^2) }\, ,
}
and the $\delta \chi$ integral in the expression for $C^{(0)}$ gives
\eeq{
\int d \delta \chi\; e^{-\delta \chi^2/(4 \sigma^2)} e^{i k_{\parallel} \delta \chi} = 2\sqrt{\pi} \sigma e^{-\sigma^2k_\pp^2}\, ,
}
which for large $\sigma$ clearly constrains the support of the integral over $k_\parallel$ to the region $|k_\pp|\lesssim1/\sigma$. 
} 
Generically, if these $\delta \chi$ window function contributions are broad 
(we can associate to them some scale $\sigma$) their Fourier transform will be 
narrow and constrained to the range $|k_\pp| \lesssim 1/\sigma$. Finally, we conclude 
that for broad enough window functions (and when the $\delta \chi$ integral bounds modes 
along the line of sight to the narrow range $|k_\pp| \lesssim 1/\sigma$), for angular modes with $\ell/\bar \chi \gg 1/\sigma$ 
we can use $P(k)\approx P(\ell/\bar\chi)$, and consequently the Limber approximation.

\subsection{Geometric recalibration}
\label{subsec:modification}

Our new derivation of the flat-sky approximate result was obtained by working in Fourier space. 
It is instructive also to tackle the problem starting from the two-point correlation function in real space. 
This approach has recently been taken in refs.~\cite{Matthewson:2020,Matthewson:2021}.
As we shall see below, this real-space analysis provides us with some insights into a geometric 
recalibration that we include in our flat-sky approximation to improve its accuracy.

\begin{figure}
    \centering
    \includegraphics[width=0.8\linewidth]{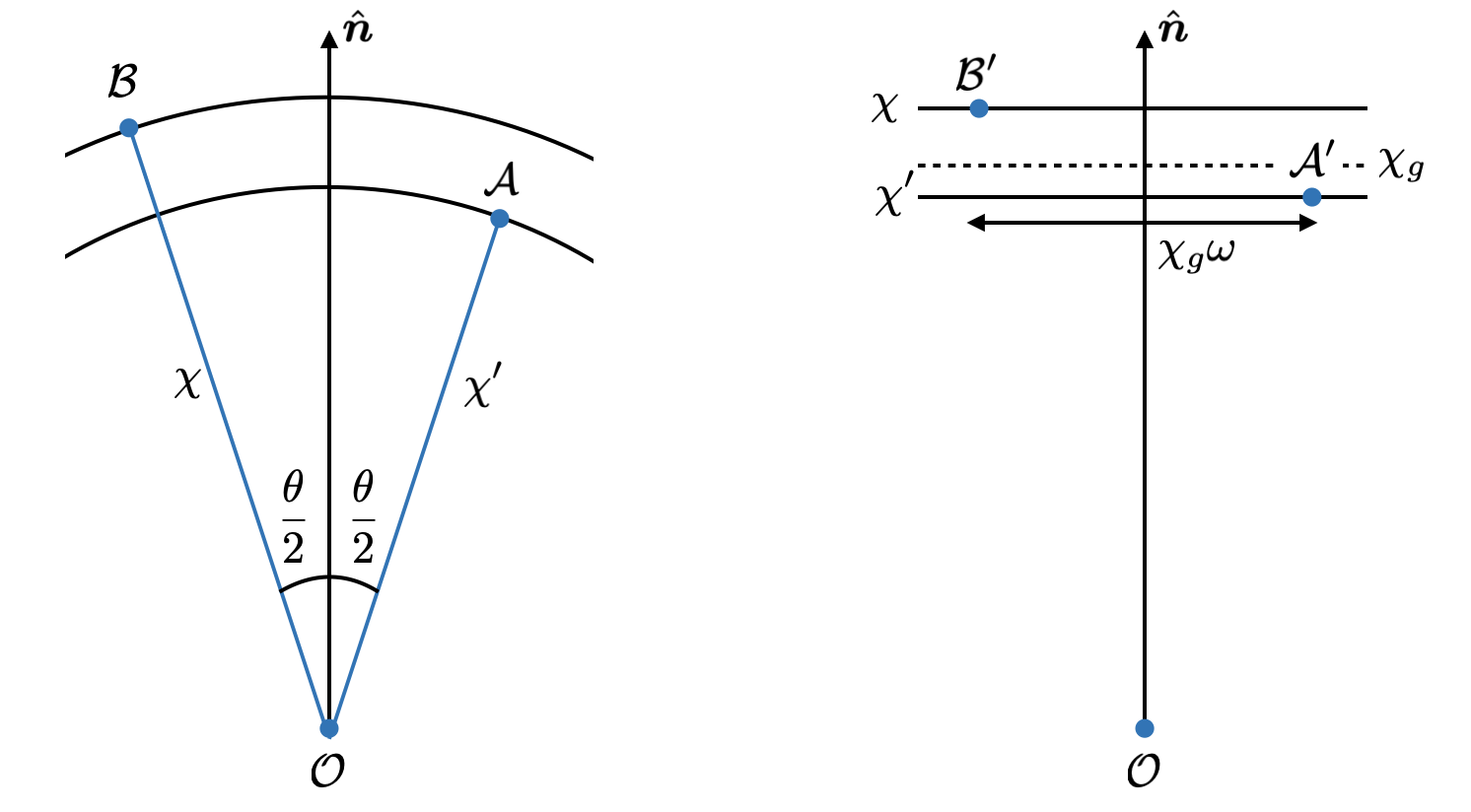}
    \caption{Geometric layout of the full-sky (left panel) and flat-sky (right panel) unequal-time, two-point 
    correlation function for points $\mathcal{A}$ and $\mathcal{B}$ separated by an angle $\theta$ and at radial distances $\chi'$ and $\chi$, respectively.
    The observer is at $\mathcal{O}$. The line-of-sight axis $\hat{\vec{n}}$ is the 
    angle bisector in the full-sky geometry. 
    Points $\mathcal{A'}$ and $\mathcal{B'}$ are corresponding points in the flat-sky set-up having the same 3D separation as $\mathcal{A}$ and $\mathcal{B}$ in the spherical case. The flat-sky angular coordinate is $\omega \equiv 2\sin(\theta/2)$ and transverse separations are determined from $\omega$ through the effective distance $\chi_g \equiv \sqrt{\chi\chi'}$.}     
    \label{Fig:Geometry}
\end{figure}

We aim to compute the unequal-time angular power spectrum $\Bbb{C}_\ell(\chi,\chi')$ from a Legendre transform of the angular correlation function:
\begin{equation}
\Bbb{C}_\ell(\chi,\chi') = 2\pi \int_{-1}^1 d\cos\theta\; \xi(\theta,\chi\,\chi') P_\ell(\cos\theta) \, .
\label{eq:xitoCell}
\end{equation}
Here, $\xi(\theta,\chi,\chi')$ is simply the 3D two-point correlation function $\xi_{\text{3D}}$ of the density contrast $\delta$ evaluated at radii $\chi$ and $\chi'$ and angular separation $\theta$. It is convenient to choose these two points $\mathcal{A}$ and $\mathcal{B}$ as in the left-hand panel of figure~\ref{Fig:Geometry}. The 3D distance between the points satisfies
\begin{align}
|\overrightarrow{\mathcal{AB}}|^2 &= (\chi+\chi')^2\sin^2(\theta/2) + (\chi-\chi')^2\cos^2(\theta/2) \nonumber \\
&= \left[\sqrt{\chi\chi'}\, 2\sin(\theta/2)\right]^2 + (\chi-\chi')^2 \, .
\end{align}
This is the same distance as between points $\mathcal{A}'$ and $\mathcal{B}'$, which lie in planes at perpendicular distances $\chi'$ and $\chi$ from the observer, respectively, and have transverse separation $\sqrt{\chi\chi'}\omega$. Here, $\omega \equiv 2 \sin(\theta/2)$ plays the role of a flat-sky angular coordinate and the effective radial coordinate $\chi_g \equiv \sqrt{\chi\chi'}$ (the geometric mean of $\chi$ and $\chi'$) is used to connect $\omega$ with transverse distances. This geometry is illustrated in the right-hand panel of figure~\ref{Fig:Geometry}. We note that the mapping from the full-sky $\theta$ to the flat-sky $\omega$ is area preserving, $d\cos\theta = \omega d\omega$. Furthermore, $\omega$ naturally arises in the asymptotic expansion of the Legenedre polynomials, as we discuss below. Expressing the 3D distance in terms of $\delta \chi$ and $\chi_g \omega$, we have
\begin{equation}
\xi(\theta,\chi,\chi') = \xi_{\text{3D}}\left(\sqrt{\chi_g^2 \omega^2 + (\delta\chi)^2}; \chi,\chi'\right) \, .
\end{equation}

The 3D correlation function is the Fourier transform of the (unequal-time) 3D power spectrum, so we can write
\begin{align}
\xi(\theta,\chi,\chi') &= \int \frac{d^3 \vec{k}}{(2\pi)^3}\; P(k;\chi,\chi') e^{i \vec{k} \cdot \vec{r}} \nonumber \\
&= \int \frac{d^3 \vec{k}}{(2\pi)^3}\; P(k;\chi,\chi') e^{i k_\parallel \delta \chi} e^{i \vec{k}_\perp \cdot (\chi_g \vec{\omega})} \, ,
\end{align}
where $\vec{\omega}$ is a 2D vector in the plane of the sky with magnitude $2\sin(\theta/2)$. We note that we have not made any approximations so far. However, to make further progress in evaluating the Legendre transform in eq.~\eqref{eq:xitoCell}, we switch the integration variable to $\omega$ and extend the domain of integration from $0\leq \omega \leq 2$ to $0\leq \omega \leq \infty$. Moreover, we approximate the Legendre polynomial by the asymptotic expansion $P_\ell(\cos\theta) \approx J_0(\sqrt{\ell(\ell+1)}\omega)$, where $J_0$ is the zeroth-order Bessel function of the first kind, valid for large $\ell$ and small $\theta$. This gives
\begin{equation}
\Bbb{C}_\ell(\chi,\chi') \approx 2\pi \int \frac{d^3 \vec{k}}{(2\pi)^3}\, P(k;\chi,\chi') e^{i k_\parallel \delta \chi} 
\int_0^\infty \omega d\omega\; J_0(\sqrt{\ell(\ell+1)}\omega) e^{i \vec{k}_\perp \cdot (\chi_g \vec{\omega})} \, .
\label{eq:flatClasymp}
\end{equation}
The integral over $\omega$ can be evaluated using the integral representation of $J_0$:
\begin{align}
2\pi \int_0^\infty \omega d\omega\, J_0(\sqrt{\ell(\ell+1)}\omega) e^{i \vec{k}_\perp \cdot (\chi_g \vec{\omega})} &= \int_0^\infty \omega d\omega \int_0^{2\pi} d\phi_{\vec{\omega}}\, e^{-i \vec{L} \cdot \vec{\omega}}  e^{i \vec{k}_\perp \cdot (\chi_g \vec{\omega})}  \nonumber \\
&= \int d^2 \vec{\omega}\; e^{i\vec{\omega} \cdot (\vec{k}_\perp \chi_g - \vec{L})} \nonumber \\
&= (2\pi)^2 \delta^{\text{2D}}(\vec{k}_\perp \chi_g - \vec{L}) \, ,
\end{align}
where $\phi_{\vec{\omega}}$ is the 2D polar angle of $\vec{\omega}$ and $\vec{L}$ is a 2D vector with magnitude $\sqrt{\ell(\ell+1)}$. Finally, evaluating the integral over $\vec{k}_\perp$ in eq.~\eqref{eq:flatClasymp}, we find
\begin{equation}
\Bbb{C}_\ell(\chi,\chi') \approx \frac{1}{\chi_g^2} \int \frac{dk_\parallel}{2\pi}\, e^{i k_\parallel \delta \chi} P\left(\sqrt{k_\parallel^2 + \frac{\ell(\ell+1)}{\chi_g^2}};\chi,\chi'\right) \, . 
\label{eq:approxcurly_sph}
\end{equation}
This can be integrated over $\chi$ and $\chi'$ [or $\delta\chi$ and $\bar{\chi}=(\chi+\chi')/2$] to obtain
\eeq{
\label{Eq:Cl_geometry}
C_\ell \approx \int \frac{d\bar{\chi} d \delta \chi}{\bar{\chi}^2(1-\delta^2)}
W\lb \bar{\chi} + \tfrac{1}{2} \delta \chi \rb W'\lb \bar{\chi} - \tfrac{1}{2} \delta \chi \rb
\int \frac{dk_{\parallel}}{2\pi} e^{i k_{\parallel}\delta\chi} P(k;\bar{\chi}, \delta\chi) \, ,
}
where $k^2 = k_\parallel^2 + \ell(\ell+1)/[\bar{\chi}^2(1-\delta^2)]$.
Equations~\eqref{eq:approxcurly_sph} and~\eqref{Eq:Cl_geometry} can be compared to our previous flat-sky results, $\Bbb{C}^{\text{flat}}(\ell)$ in eq.~\eqref{Eq:curly_flat} and $C^{\text{flat}}(\ell) = C^{(0)}(\ell)$ in eq.~\eqref{Eq:Cln}.
The only differences are the replacement $\ell \rightarrow \sqrt{\ell(\ell+1)}$ and the presence of $\chi_g^2 = \bar{\chi}^2(1-\delta^2)$, rather than simply $\bar{\chi}^2$, in the prefactor in eq.~\eqref{eq:approxcurly_sph}. The replacement $\ell \rightarrow \sqrt{\ell(\ell+1)} \approx \ell + 1/2$ is widely used in applications of the Limber approximation. We refer to it as \emph{geometric recalibration}. As we show in section~\ref{Sec:RD}, we find universally better agreement between our flat-sky results and their full-sky counterparts when including this geometric recalibration. For the results in this paper, we use the prefactor $1/\bar{\chi}^2$ rather than $1/\chi_g^2$ as both give very similar results.

\subsection{Evaluation of the flat-sky angular power spectrum}
\label{subsec:evaluation}

In this subsection we discuss our strategy for numerical evaluation of the 
flat-sky angular power spectrum given in, for example, eqs.~\eqref{Eq:Cln} and \eqref{Eq:curly_flat}.
The integral over $k_\pp$ is a one-dimensional Fourier transform, which 
makes the evaluation fairly straightforward.
Nonetheless, direct evaluation using the discrete Fourier transform can still be somewhat time 
consuming. This is especially so when the residual $\delta \chi$ dependencies in the 3D power 
spectrum are kept since then one cannot use a single Fourier transform to obtain a grid of 
corresponding $\delta \chi$s, but rather a single transform for each $\delta \chi$ point is required.

In this paper we take a different approach, relying on the fact that the 3D power spectra (be it linear 
or nonlinear) can be well represented by the discrete Mellin transform. The pioneering application 
of this method in cosmology was fast-Fourier-transforming the 3D power spectrum and 
correlation function in ref.~\cite{Hamilton2000}; the algorithm has been named FFTLog.
Since then, FFTLog has also been used in the computation of nonlinear corrections 
to cosmological correlators~\cite{Schmittfull:2016, McEwen:2016, Schmittfull:2016_II, Simonovic:2017, Tomlinson:2020}, 
as well as in the evaluation of the full-sky angular power spectrum~\cite{Assassi2017,Gebhardt:2017,Schoneberg:2018,Fang:2020}.
We will follow a similar route to ref.~\cite{Assassi2017}, but with the difference of applying the FFTLog algorithm
to compute the flat-sky angular power spectrum. The advantage, as we shall see, is that the resulting flat-sky expressions are significantly simpler than their full-sky counterparts, yielding a 
significant computationally speed-up.

Our starting point is thus to represent the 3D power spectrum 
$P(k)$ in terms of a sum of (complex) powers of the wavenumber $k$,
i.e., $P(k) \simeq \sum_{i} \alpha_i k^{\nu_i}$. In this work, we constrain our analysis 
to the linear version of the 3D power spectrum for which the time dependence is separable:
\eeq{
\label{Eq:P_linear}
P \left(k;\, \bar{\chi}, \delta \chi \right) = 
D\lb \bar{\chi}+\tfrac{1}{2}\delta\chi\rb D\lb \bar{\chi}-\tfrac{1}{2}\delta\chi\rb p(k)\, ,
}
where $p(k)$ is the 3D linear power spectrum at $z=0$ and $D(\chi)$ is the linear growth factor normalised to unity at $z=0$.\footnote{\ZG{For simplicity, we have assumed a linear 3D power spectrum. We can also include the nonlinear corrections, as well as the unequal-time effects (see, e.g. \cite{Raccanelli_II:2023}). The only significant change this introduces is the possible explicit time dependence in the $\alpha_i$ coefficients.}}
Working with the linear form of the power spectrum provides certain simplifications,
however, it is important to stress that in no significant way does this represent a limitation of the
method, and a similar procedure can be adopted when using any nonlinear $P(k)$ results.
As discussed earlier, the relation between multipoles $\ell$ and 3D wavenumbers $k$ is, after geometric recalibration,
\eeq{
\label{Eq:lk}
k = \sqrt{k_{\parallel}^2+\tilde \ell^2} = \sqrt{k_{\parallel}^2+\frac{\ell(\ell+1)}{\bar{\chi}^2 (1-\delta^2)}}\, ,
} 
where we introduced the shorthand notation $\tilde \ell \equiv \sqrt{\ell(\ell+1)}/(\bar{\chi} \sqrt{1-\delta^2})$. We note that $\tilde{\ell}$ has dimensions of inverse length.
We use the FFTLog algorithm to obtain the coefficients and powers of the expansion 
$p(k) = \sum_{i} \alpha_i k^{\nu_i}$, where we note that the frequencies $\nu_i$ are complex numbers with fixed negative real part (known as the bias). One can  
always change the real bias term to the frequencies, with an associated change in the coefficients $\alpha_i$, by multiplying $p(k)$ by the appropriate power of $k$ before performing the transform. In our implementation, we have found that using $\Re(\nu_i) = -0.5$ gives excellent recovery of the 3D power spectrum, however, one can always change the biasing term in our code, if desired.

The flat-sky result for the unequal-time angular power spectrum given 
in eq.~\eqref{Eq:curly_flat} can thus be written as 
\eeq{ 
\label{Eq:Expansion} 
\mathbb C^{\text{flat}}(\ell, \bar{\chi}, \delta\chi) = \frac{1}{\bar \chi^2} \sum_i \alpha'_i
 \int_{-\infty}^\infty \frac{d k_{\parallel}}{2\pi} ~ e^{ i  k_{\parallel} \delta \chi}
\lb k_{\parallel}^2+{\tilde\ell}^2 \rb ^{\frac{\nu_i}{2}}
= \frac{1}{\bar \chi^2} \sum_i \alpha'_i\, \frac{(2\tilde{\ell}/\delta\chi)^{\frac{1}{2}(\nu_i+1)}}{\sqrt{\pi}\Gamma(-\frac{\nu_i}{2})} 
K_{\frac{1}{2}(\nu_i+1)}(\tilde{\ell}\delta\chi)\, ,
}
where $K_\nu(z)$ is the modified Bessel function of the second kind, 
and where we absorbed the $D\lb \bar{\chi}+\delta\chi/2\rb D\lb \bar{\chi}-\delta\chi/2\rb$ factor in the $\alpha'_i$ coefficients.
It is useful to introduce a function defined as
\eq{
\label{Eq:M2i}
M^{(2)}_{\nu}(x) \equiv 
\frac{2^{\frac{1}{2}(\nu+1)}}{\sqrt{\pi}\Gamma(-\frac{\nu}{2})} x^{-\frac{1}{2}(\nu+1)
} K_{\frac{1}{2}(\nu+1)}(x)\, ,
}
which enables us to rewrite the unequal-time angular power spectrum  $\mathbb{C}^{\text{flat}}(\ell)$ in the following form:
\eeq{
\label{Eq:SpecialFunction}
\mathbb{C}^{\text{flat}}(\ell, \chi, \delta\chi) 
= \frac{1}{\bar \chi^2} \sum_i \alpha'_i \, \tilde{\ell}^{\nu_i+1} M^{(2)}_{\nu_i}\big( \tilde{\ell}\delta\chi \big)\,.
}
We note that $M^{(2)}_{\nu}(x)$ is an even function,\footnote{We can use the property $I_\nu(-z)=(-1)^{\nu}I_{\nu}(z)$, and since $K_\nu(z)=\pi[ I_{-\nu}(z)-I_\nu(z)] / [2\sin(\nu \pi)]$, it follows that $K_{\nu}(-z)=(-1)^\nu K_\nu(z)$.} 
i.e., $M^{(2)}_{\nu}(-x)=M^{(2)}_{\nu}(x)$, which is a
property inherited from the fact that $\mathbb{C}^{\text{flat}}(\ell, \bar{\chi}, \delta\chi) = \mathbb{C}^{\text{flat}}(\ell, \bar{\chi}, -\delta\chi)$.
The representation of our results in terms of the $M^{(2)}_{\nu}(x)$ functions provides us with several benefits. 
First, we note that these functions do not carry any information on the 
cosmological parameters, i.e., in any cosmological analysis (and once the grid of $\nu_i$
has been fixed) $M^{(2)}_{\nu_i}(x)$ can be pre-computed and interpolated in the $x$ variable. 
This allows for almost instantaneous evaluation of these functions for any choice of $x = \tilde{\ell}\delta\chi$.
Note that the argument $\tilde{\ell}\delta\chi$ depends on cosmology via the definition of the radial distances; however, we can ensure that $M^{(2)}_{\nu_i}(x)$ is pre-computed over a sufficiently wide range to encompass any reasonable variation in cosmology.
The rest of the cosmological information is, of course, carried
by the $\alpha'_i$ coefficients obtained by the single FFTLog decomposition of the 
linear power spectrum. The second beneficial property is that we require (pre-evaluation) of only $N_\nu$ special functions 
$M^{(2)}_{\nu_i}(x)$ functions, where $N_\nu$ is the number of frequencies used in the FFTLog expansion of $p(k)$.
This is to be contrasted with the evaluation of the full-sky $\mathbb{C}_\ell^{\rm full}$ with the FFTLog expansion, as 
put forward in refs.~\cite{Assassi2017,Schoneberg:2018}
and which we summarise in Appendix~\ref{app:full_sky_Cell}. The equivalent full-sky 
representation of the angular power spectrum is given in eq.~\eqref{eq:full_sky_cell};
it depends on the functions $I_{\ell}\lb \nu, t \rb$ (defined in eq.~\ref{Eq:I_l1_l2} and where $t=\chi'/\chi$).
For a given $\nu_i$, separate evaluations of special functions (in this case, involving the hypergeometric function ${}_2 F_1$) are required over a grid of $\ell$ and $t$ values as the arguments do not combine into the single $x = \tilde{\ell}\delta \chi$ as in the flat-sky case.
The full-sky calculation thus requires a much larger number of special-function evaluations. Moreover, in each evaluation 
the hypergeometric function ${}_2 F_1$ is required, which is more involved and 
thus computationally more costly than the $M^{(2)}_{\nu}(x)$. Being able to pre-compute the functions $M^{(2)}_{\nu_i}(x)$ and then interpolate the argument makes the computational performance advantage of the flat-sky approximation even starker. 

In our default setup, we use $N_\nu=201$ modes for the expansion of $p(k)$, 
which is sufficient to reproduce precisely the linear power spectrum at $z=0$. We typically choose $k_{\text{min}}=10^{-8}\, h/\mathrm{Mpc}$ and $k_{\text{max}}=52.0\, h/\mathrm{Mpc}$ when taking the transform of $p(k)$. With these choices, the fractional error in the reconstructed $p(k)$ is below $0.5\%$ for $5\times 10^{-4}\, h/\mathrm{Mpc}\leq k \leq 8\, h/\mathrm{Mpc}$. One can further optimise the choice of frequencies if required. As mentioned, we then pre-calculate the set of $M^{(2)}_{\nu_i}(x)$ functions
choosing 2000 sampling points for their $x$ arguments. For projected galaxy clustering, if we sample
$\bar{\chi}$ and $\delta\chi$ at 50 and 100 points, respectively (which we found to provide a converged result), it takes about 
$12\,\text{s}$ to calculate 150 multipoles in Python3 on a personal laptop.
Note that these timings do not include generation of the arrays of window functions and growth factors, and depend on the choice of the length of sampling arrays. 
The algorithm could certainly be further optimised for application in likelihood analysis, where computational time is critical. The details of our calculation and code documentation can be found in the GitHub repository.\footnote{https://github.com/GZCPhysics/BeyondLimber.git}

\section{Results for galaxy clustering and CMB lensing}
\label{Sec:RD}

In this section, we first use our results to compute the projected angular power spectrum in the case of galaxy clustering. We then extend this analysis by including redshift-space distortions. Lastly, we turn to the case of CMB lensing. We compare our results to the full-sky FFTLog-based calculation following ref.~\cite{Assassi2017}, which we have also implemented in our code. In each case, we also show comparisons with the Limber approximation.

\subsection{Galaxy clustering}
\label{Sec:GalaxyClustering}

We focus first on projected galaxy clustering, where galaxies are selected 
within a given redshift range according to the radial window function $W(\chi)$. 
We assume a simple case where the window function is
Gaussian, specified by a central distance $\chi_*$ and width $\sigma_\chi$, i.e.,
\eeq{
W_g(\chi) = \frac{1}{\sqrt{2\pi} \sigma_\chi} \exp\left[ -\frac{(\chi-\chi_*)^2}{2\sigma_\chi^2}\right]\, ,
}
where, for narrow window functions, $\sigma_\chi$ is related to the standard deviation of redshift via $\sigma_\chi = c \sigma_z /H(z)$ for $a_0=1$. We highlight that our computation of the
projected angular power spectrum $C_\ell$, as given in eq.~\eqref{Eq:Cl_geometry},
is not particularly sensitive to characteristics of $W(\chi)$ such as its smoothness. In particular, 
we do not require $W(\chi)$ to be differentiable (a property used, for example, in \cite{Assassi2017})
or to have a smooth and well-behaved Fourier transform (utilised in \cite{Fang:2020}).
This may prove helpful in analyses of survey data, where the window functions may be complicated due to photometric redshift uncertainties, for example. Rather, 
we focus on developing fast computational methods for the unequal-time angular power spectrum $\mathbb C_\ell$, which is then integrated against the pair of window functions, without needing to put additional constraints on these, to obtain the projected spectrum $C_\ell$.

\begin{figure}[t!]
\centering
\begin{subfigure}[b]{0.48\textwidth}
\includegraphics[width=\linewidth]{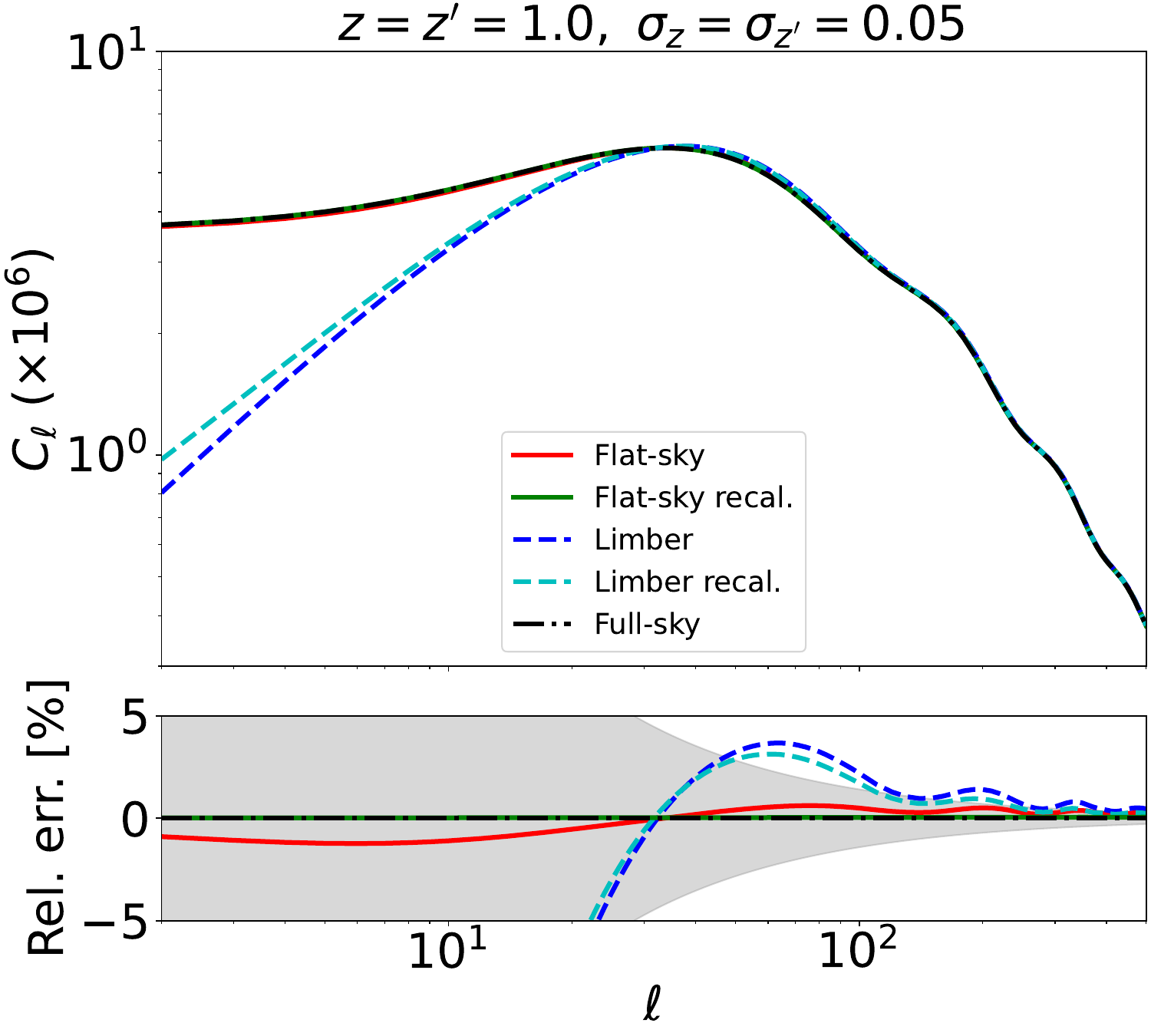}
\end{subfigure}
\begin{subfigure}[b]{0.48\textwidth}
\includegraphics[width=\linewidth]{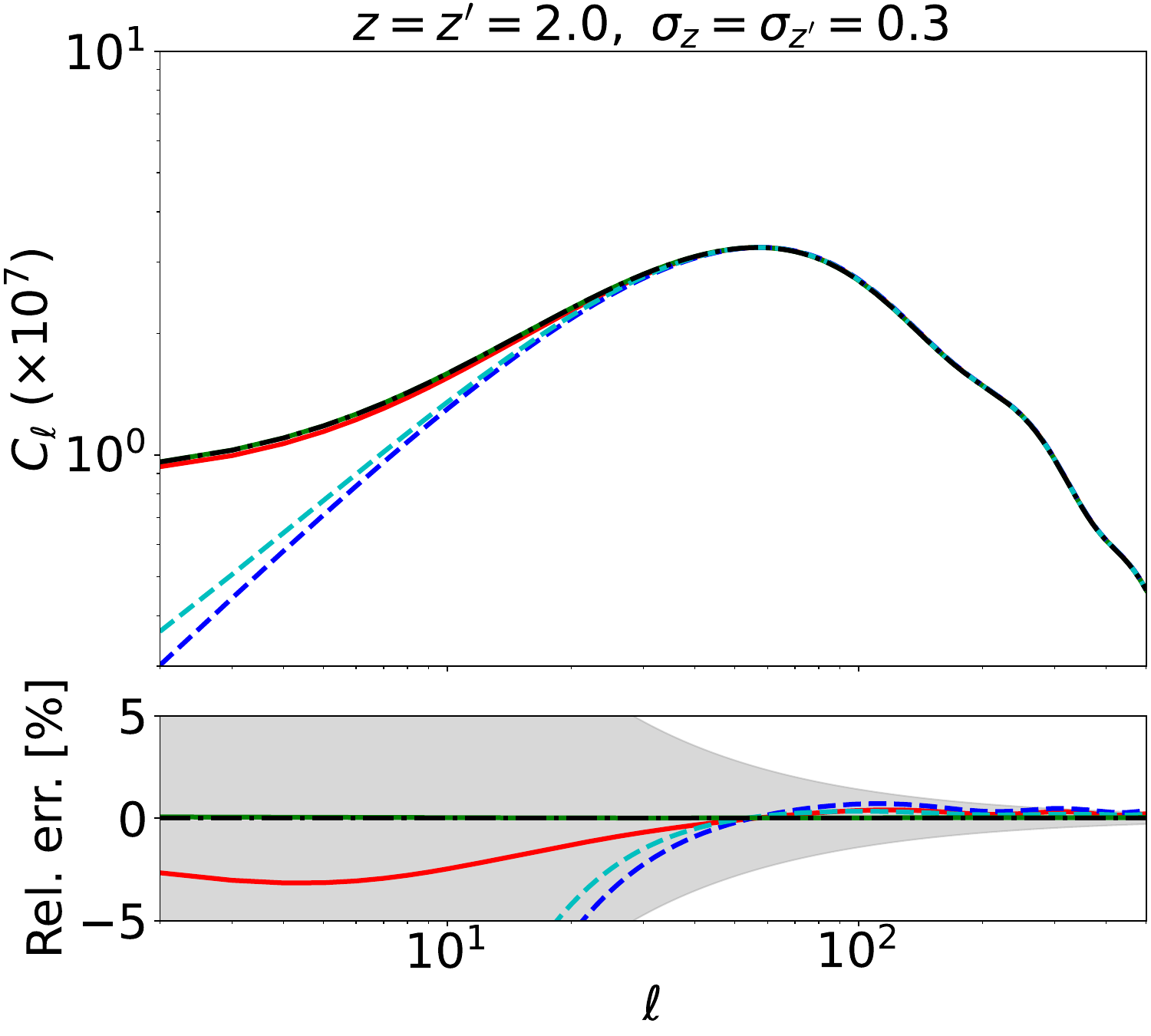}
\end{subfigure}
\caption{Angular power spectra of projected galaxy clustering with equal-time window functions. 
The left panel is for central redshifts $z=z'=1.0$ and width $\sigma_z = 0.05$, 
while the right panel is for $z=z'=2.0$, with $\sigma_z = 0.3$. 
In both panels, we compare the full-sky expression (black dash-dotted line), 
the Limber approximation (blue dashed line), and the flat-sky approximation (red solid line). 
The green solid line and cyan dashed line are geometrically recalibrated flat-sky and Limber approximations, 
as described in section~\ref{subsec:modification}. 
In the bottom panels we show fractional residuals compared to the full-sky results. \ZG{The grey bands represent the (fractional) statistical error in the amplitude of the power spectrum, using all multipoles less than $\ell$, in the cosmic-variance limit as described in the text.}}
\label{fig:Cell_galaxy_clustering_auto}
\end{figure}

In figure~\ref{fig:Cell_galaxy_clustering_auto}, we present results for the angular power spectrum of galaxy clustering projected 
with equal window functions centred on redshifts $z=1.0$ and $2.0$,
and with widths $\sigma_z = 0.05$ and  $\sigma_z = 0.3$, respectively.
We find excellent overall agreement of the full- and flat-sky results, on all scales,
while the Limber approximation deviates significantly from these on large scales. 
For multipoles $\ell\lesssim50$, the flat-sky approximation (before geometric recalibration) only deviates from the 
target full-sky angular power spectrum by a few percent at most. These small differences between the full- and flat-sky results are removed very effectively when the additional geometric recalibration $\ell \to \sqrt{\ell(\ell+1)}$
is implemented in the flat-sky result (see the discussion in section~\ref{subsec:modification}).
Geometric recalibration in the Limber approximation also improves its agreement with the full-sky result somewhat, but starting from much larger errors than our flat-sky results.
We note that for the relatively narrow window functions (left panel of figure~\ref{fig:Cell_galaxy_clustering_auto}), the fractional error in the Limber approximation remains at the percent level or higher up to multipoles of a few hundred.
The Limber approximation performs better for the higher-redshift and broader window functions (the right panel of figure~\ref{fig:Cell_galaxy_clustering_auto}), as expected from the discussion in section~\ref{subsec:limber}.

\ZG{When considering what size of theoretical errors in the power spectrum can be tolerated, it is not sufficient to compare to the cosmic-variance error per multipole since many multipoles are combined to estimate cosmological parameters and the theoretical errors may be coherent across the multipole range. Instead, we adopt the following procedure here. Consider estimating an amplitude parameter $A$ that simply scales a fiducial spectrum, so the true value is $A=1$. If we use all multipoles less than $\ell_\text{max}$ when estimating $A$, then its cosmic-variance error is $\sqrt{2}/\ell_\text{max}$. If the fractional theoretical error on the power spectrum $\epsilon_\ell$ is less than this cosmic-variance error for all $\ell < \ell_\text{max}$, the bias in $A$ is ensured to be less than cosmic variance. We show $\sqrt{2}/\ell$ by the grey band in figure~\ref{fig:Cell_galaxy_clustering_auto}.}

\begin{figure}[t!]
\centering
\begin{subfigure}[b]{0.48\textwidth}
\includegraphics[width=\linewidth]{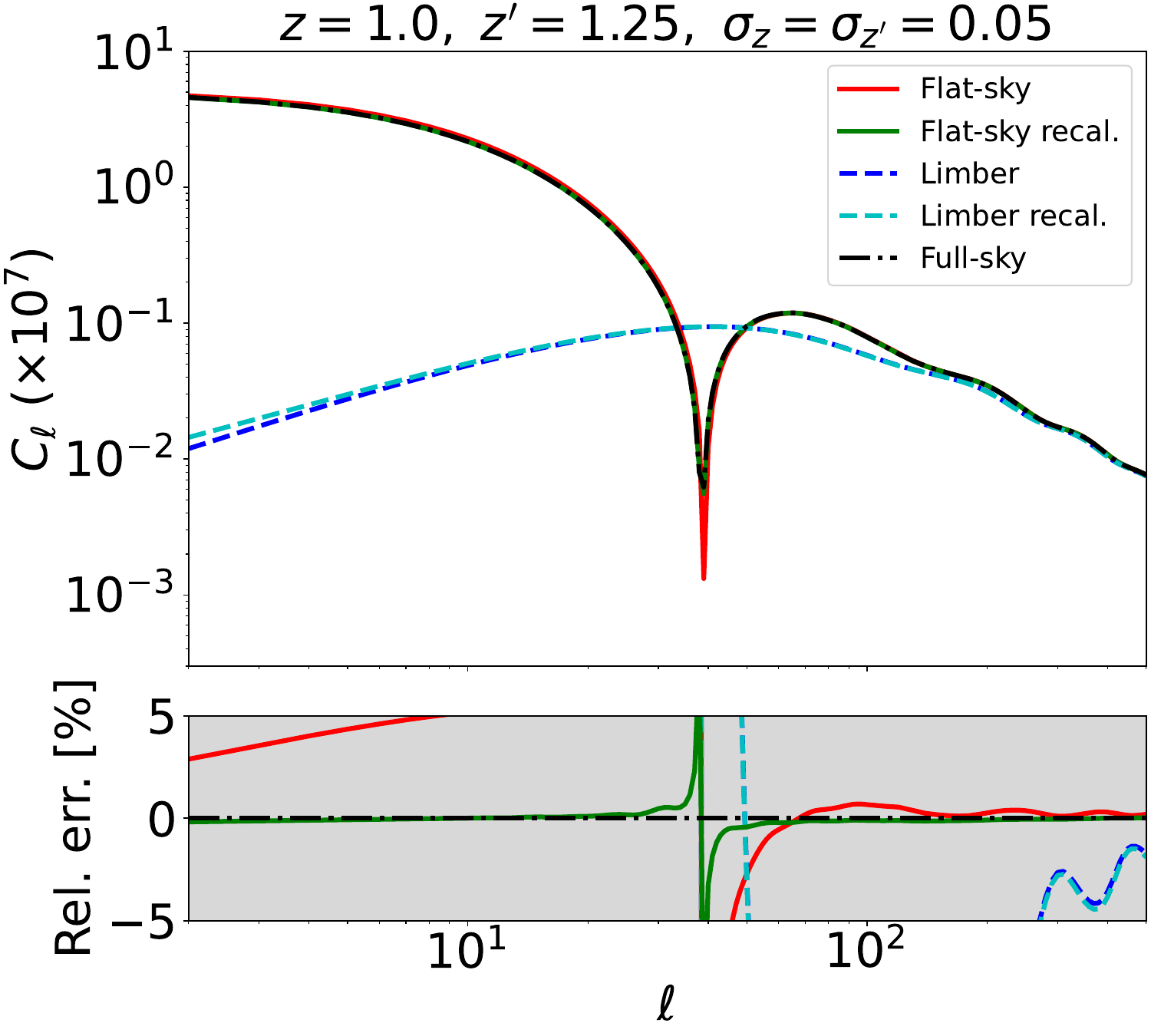}
\end{subfigure}
\hspace{0.1cm}
\begin{subfigure}[b]{0.48\textwidth}
\includegraphics[width=\linewidth]{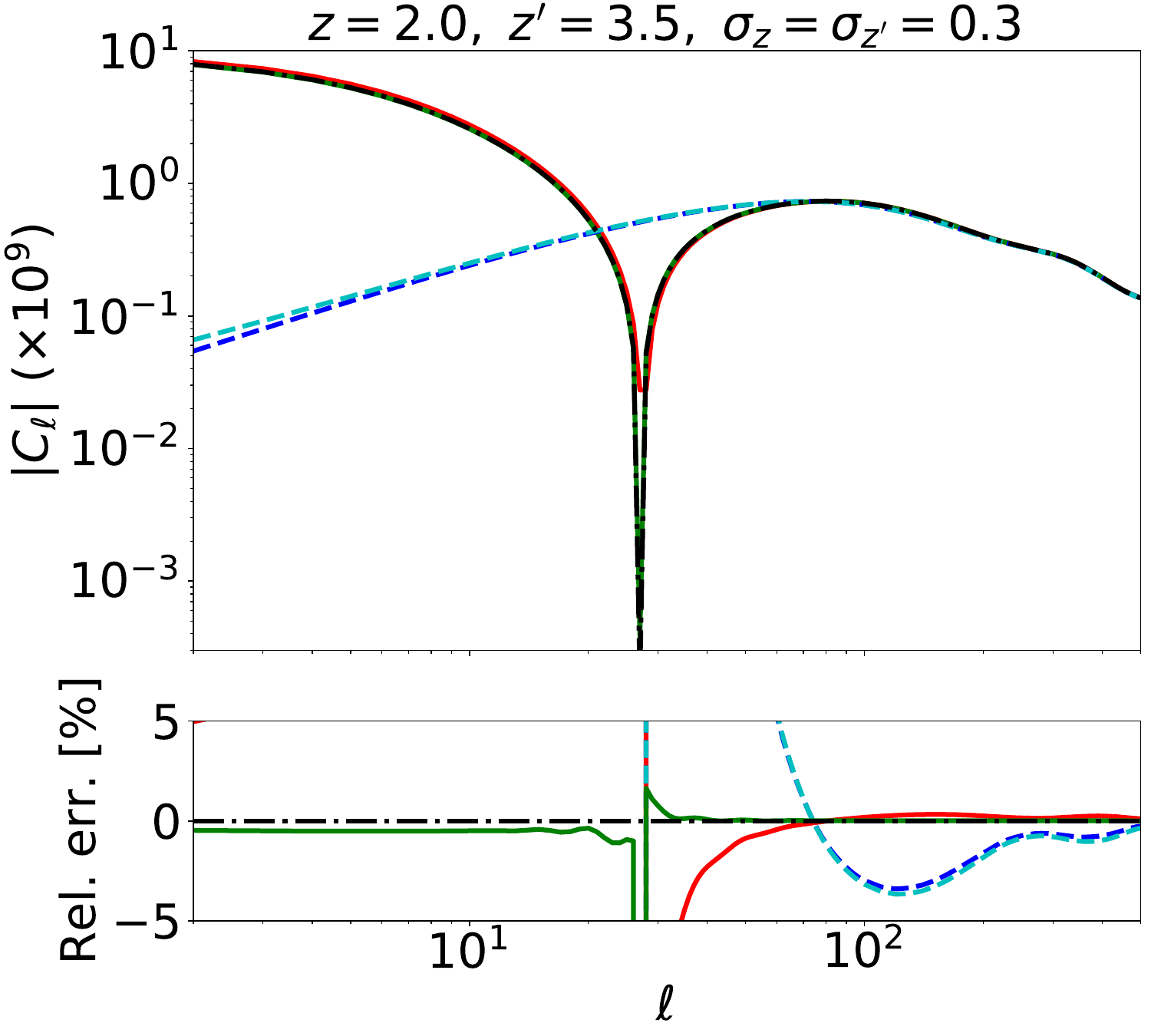}
\end{subfigure}
\caption{As figure~\ref{fig:Cell_galaxy_clustering_auto} but for unequal-time window functions. 
The left panel is for central redshifts $z=1.0$, $z'=1.25$, with $\sigma_z = 0.05$ in both cases, 
while the right panel is for $z=2.0$, $z'=3.5$, with $\sigma_z = 0.3$. The true angular power spectrum is negative on large scales (the absolute value is plotted).}
\label{fig:Cell_galaxy_clustering_cross}
\end{figure}

We now consider the cross-correlation between the signal projected with unequal-time window functions. The magnitude of the  angular power spectrum is shown in figure~\ref{fig:Cell_galaxy_clustering_cross} for two setups: first with central redshifts $z=1.0$, $z'=1.25$ and equal widths $\sigma_z = 0.05$; and second with 
$z=2.0$, $z'=3.5$ and $\sigma_z = 0.3$. We again find very good
overall agreement of the full- and flat-sky results, especially once the 
geometric recalibration is taken into account. In both scenarios considered, the angular power spectrum 
is negative (anti-correlated) on large scales while remaining positive on smaller scales. 
This anti-correlation feature is also captured very well by the flat-sky 
result, providing us again with sub-percent agreement with the full-sky result 
on all scales (away from the zero crossing) after geometric recalibration. We highlight that even without the recalibration,
the flat-sky result correctly captures the sign change and the shape of the full-sky results. 
The effect that the $\ell \to \sqrt{\ell(\ell+1)}$ recalibration provides is a slight horizontal shift to lower multipoles on large scales. 
In contrast, the Limber approximation fails to capture the sign change on large scales, being strictly positive on 
all scales, and, in general, performs poorly for cross-correlations with little overlap between the radial window functions. 

As mentioned in section~\ref{sec:ang_ps}, the key difference between the Limber approximation 
and our flat-sky result is that the latter retains information about the $k_\pp$ wave-modes, which the Limber approximation
explicitly disregards. As a result, the flat-sky approximation can model both equal-time and unequal-time correlations. 
The inclusion of these modes along the line-of-sight is essential to capture the correct behaviour of the cross-power spectrum for unequal-time window functions on large scales, where they lead to anti-correlations (figure~\ref{fig:Cell_galaxy_clustering_cross}).
Keeping the integration over $k_\pp$ is thus the primary source of the improvements we observe 
in the flat-sky results over those from the Limber approximation. Both of these approximations, of course, fail to account 
for the effects of sky curvature beyond what is already captured by the geometric recalibration
$\ell \to \sqrt{\ell(\ell+1)}$. However, our results suggest that these are only relatively minor corrections 
for any realistic clustering survey setup and geometry, even on the largest scales ($\ell < 10$). 

\begin{figure}
    \centering
    \includegraphics[width = 0.6\textwidth]{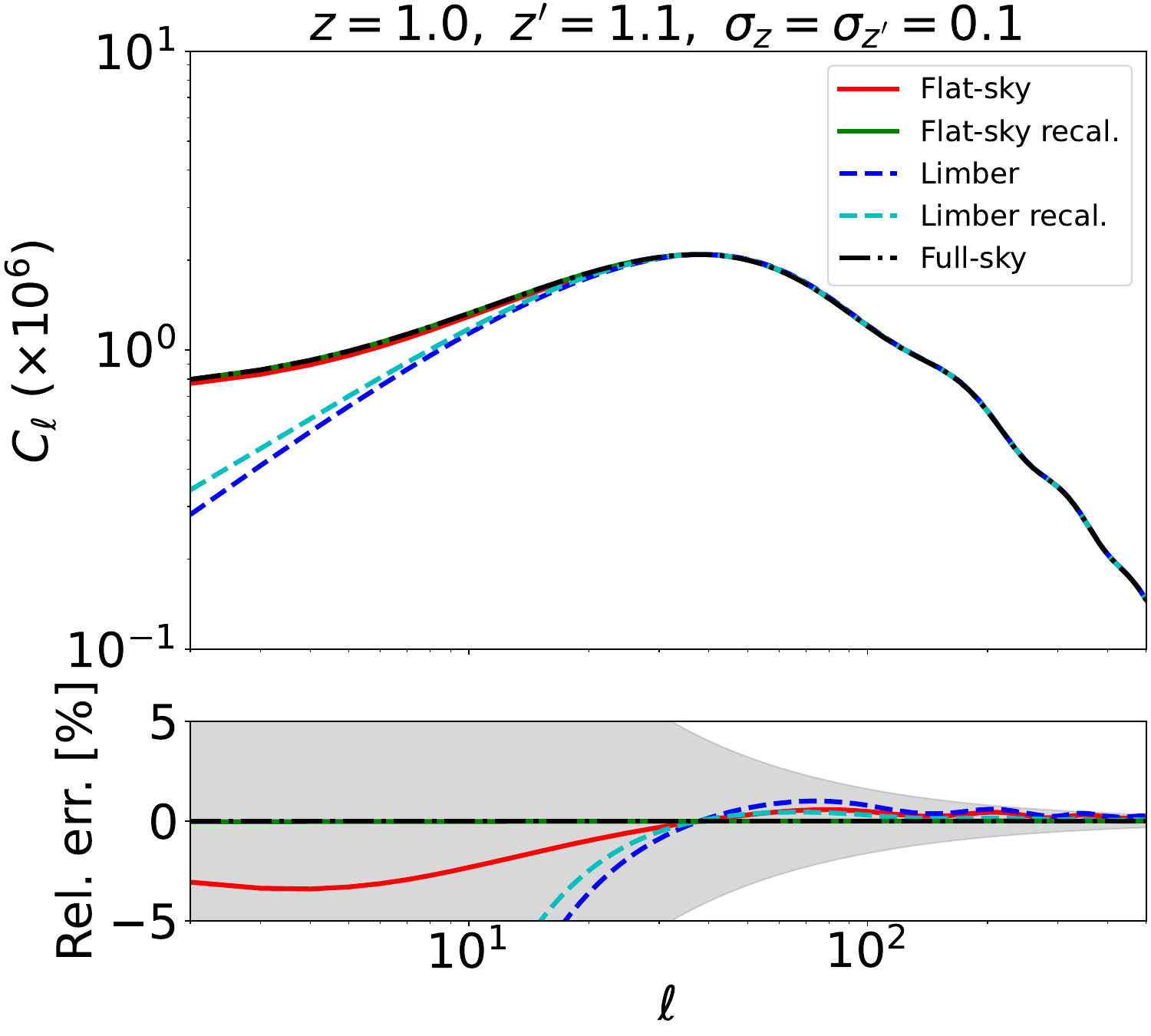}
    \caption{Angular power spectrum for the cross-correlation of galaxy clustering with central redshifts $z=1.0$, $z'=1.1$ and equal widths $\sigma_z = 0.1$. \ZG{The grey band is derived from the cosmic variance on the amplitude of the cross power spectrum; see eq.~\eqref{eq:crosserrorbound}.}}
    \label{Fig:z111}
\end{figure}

\ZG{To assess the accuracy required in a cross-spectrum, we note that if the maximum fractional error in $C_\ell^{XY}$ for $\ell < \ell_\text{max}$ satisfies
\begin{equation}
\text{max}(\epsilon_\ell) <     \left( \sum_{\ell=2}^{\ell_\text{max}} \frac{(2\ell+1)}{1+1/r_{\ell}^2}\right)^{-1/2} \, ,
\label{eq:crosserrorbound}
\end{equation}
the bias in the amplitude of the cross-spectrum estimated from multipoles less than $\ell_\text{max}$ will be less than the cosmic variance error.
Here, the correlation coefficient $r_\ell \equiv C_\ell^{XY}/\sqrt{C_\ell^{XX}C_\ell^{YY}}$. For widely separated radial window functions, the correlation is very low and the cosmic-variance error is dominated by chance fluctuations in the two observables $X$ and $Y$. In this limit, the requirement on the fractional accuracy of the cross-spectrum is very weak and too large to show in figure~\ref{fig:Cell_galaxy_clustering_cross}. For this reason, we show in figure~\ref{Fig:z111} a case with much larger correlation (central redshifts $z=1.0$ and $z'=1.1$ and widths $\sigma_z = \sigma_{z'} = 0.1$) and include the right-hand side of eq.~\eqref{eq:crosserrorbound} as a grey band. We see again the excellent accuracy of the geometrically recalibrated flat-sky result.}

\subsection{Galaxy clustering with redshift-space distortions}
\label{subsec:RSD}

In addition to the Hubble expansion, galaxies experience an additional, 
large-scale peculiar motion due to gravitational infall . This motion, 
via the Doppler effect, causes the apparent displacement of the real-space galaxy 
distribution along the line-of-sight direction when observed in redshift.
In linear theory (assuming the flat-sky and distant-observer approximation), the effect on the redshift-space over-density of sources (e.g., galaxies) can be expressed as 
\eeq{
\delta_s(\vec{k}, z) = \lb b(k,z) + f(z) k^2_\pp/k^2 \rb \delta(\vec{k}, z) \, ,
\label{eq:kaiser}
}
where $f \equiv d \ln D/d \ln a$ is the logarithmic growth rate, 
and $b$ is the linear galaxy bias, which we set to unity throughout this paper.
Equation~\eqref{eq:kaiser} is the well-known Kaiser effect~\cite{Kaiser:1987, Hamilton:1997} describing redshift-space distortions (RSD). 
The unequal-time 3D linear power spectrum of $\delta_s$ is then simply given as 
\eeq{
\label{Eq:Pk_RSD}
P_s(\vec{k};\,\chi,\chi') = \lb 1+f(\chi) k^2_{\parallel}/k^2\rb \lb 1+f(\chi') k^2_{\parallel}/k^2\rb P_L(k;\,\chi,\chi')\, ,
}
where $P_L(k;\,\chi,\chi')$ is the 3D linear matter power spectrum at lookback times $\chi$ and $\chi'$. Note that $P_s(\vec{k})$ depends separately on $k_\parallel$ and $k_\perp$ as the fixed line-of-sight breaks statistical isotropy.
Using the 3D RSD power spectrum in the expression for the unequal-time angular power spectrum
(eq.~\ref{Eq:curly_flat}), and keeping track separately of $k_\parallel$ and $k_\perp$, we have
\eeq{
\label{Eq:curly_flat_RSD}
\mathbb C^{\text{flat}}(\ell,\bar{\chi}, \delta \chi) = \frac{1}{\bar \chi^2}
\int \frac{d k_{\parallel}}{2\pi} ~ e^{ i k_{\parallel}\delta \chi} 
\lb 1+\ls f(\chi)+f(\chi') \rs \frac{k_{\parallel}^2}{k^2} + f(\chi)f(\chi')\frac{k_{\parallel}^4}{k^4}\rb
P_L \left( k;\, \bar{\chi}, \delta \chi \right)\, ,
}
where $k = \sqrt{k_{\parallel}^2 + \tilde{\ell}^2}$.
We decompose the linear power spectrum as a sum of power laws in the same way as in section~\ref{subsec:evaluation}.
To handle the additional $k_{\parallel}^2$ and $k_{\parallel}^4$ terms, 
we take derivatives of the expression given in eq.~\eqref{Eq:Expansion} with respect to $\delta\chi$ (treating $\tilde{\ell}$ and $\bar{\chi}$ as parameters that are not differentiated) and simultaneously shift the $\nu_i$ index down.
We obtain the following:
\eq{
\int_{-\infty}^\infty \frac{d k_{\parallel}}{2\pi} \; e^{ i \delta \chi k_{\parallel}} k_{\parallel}^2
\lb k_{\parallel}^2+{\tilde\ell}^2 \rb ^{\nu_i/2-1}
&= - \tilde{\ell}^{\nu_i+1} \left. \frac{d^2 M^{(2)}_{\nu_i -2}}{dx^2} \right|_{x = \tilde{\ell}\delta \chi} \, ,\\
\int_{-\infty}^\infty \frac{d k_{\parallel}}{2\pi} \; e^{ i \delta \chi k_{\parallel}} k_{\parallel}^4
\lb k_{\parallel}^2+{\tilde\ell}^2 \rb ^{\nu_i/2-2}
&= \tilde{\ell}^{\nu_i+1} \left. \frac{d^4 M^{(2)}_{\nu_i -4}}{dx^4} \right|_{x = \tilde{\ell}\delta \chi} \, ,
}
where the functions $M^{(2)}_\nu(x)$ are defined in eq.~\eqref{Eq:M2i}. Evaluating the derivatives using relations for derivatives of modified Bessel functions, we have
\eq{
M^{(3)}_\nu(x) &\equiv -\frac{d^2 M^{(2)}_\nu}{dx^2} = -\frac{2^{\frac{1}{2} (\nu+1)}}{\sqrt{\pi} \Gamma\lb-\frac{\nu}{2}\rb} x^{-\frac{1}{2}(\nu+1)} \ls K_{\frac{1}{2}(\nu+1)}(x) + (\nu+2)x^{-1} K_{\frac{1}{2}(\nu+3)}(x)\rs \, , \label{Eq:M3} \\
M^{(4)}_\nu(x) &\equiv \frac{d^4 M^{(2)}_\nu}{dx^4} = \frac{2^{\frac{1}{2} (\nu+1)}}{\sqrt{\pi} \Gamma\lb-\frac{\nu}{2}\rb} x^{-\frac{1}{2}(\nu+1)} \left[ K_{\frac{1}{2}(\nu+1)}(x) +2(\nu+2)x^{-1}K_{\frac{1}{2}(\nu+3)}(x) \right. \nonumber \\
&\hspace{0.45\textwidth} \left. + (\nu+2)(\nu+4)x^{-2}K_{\frac{1}{2}(\nu+5)}(x) \right] \, . \label{Eq:M4}
}
The treatment and numerical implementation of these functions is similar to our discussion of $M^{(2)}_\nu(x)$
in section~\ref{subsec:evaluation}.

Using the FFTLog expansion, eq.~(\ref{Eq:curly_flat_RSD}) can now be written as (where the linear growth factors are again absorbed into the $\alpha'_i$ coefficients)
\begin{multline}
\label{Eq:curly_flat_RSD_II}
\mathbb C^{\text{flat}}(\ell,\chi, \delta \chi) = \frac{1}{\bar \chi^2}\sum_i \alpha'_i \tilde{\ell}^{\nu_i+1} \left( M^{(2)}_{\nu_i}(\tilde{\ell}\delta\chi)
 + \ls f(\chi)+f(\chi') \rs M^{(3)}_{\nu_i-2}(\tilde{\ell}\delta\chi) \right. \\
 \left. + f(\chi)f(\chi')M^{(4)}_{\nu_i-4}(\tilde{\ell}\delta\chi)\right) \, .
\end{multline}
This is the expression we have implemented in our code.
Implementation requires two additional sets of pre-calculated special functions compared to the non-RSD case. 
As a result, the total pre-calculation time increases approximately three-fold (taking around $15\,\text{min}$ on a personal laptop, using 2000 sampling points for $\tilde{\ell} |\delta\chi|$). For a given choice of window functions and their sampling,
the computational time for $C^{\text{flat}}(\ell)$ including RSD also increases five-fold to approximately $0.45\,\text{sec}$ per multipole.

Before presenting numerical results, let us briefly discuss the Limber approximation in the presence of RSD.
In its strictest form, RSD do not contribute in the Limber approximation since they require non-zero $k_{\parallel}$.
However, by relaxing the Limber assumptions to a certain level, ref.~\cite{Tanidis:2019} provides 
a way of extending the Limber approximation to include RSD.\footnote[2]{%
This extension can be implemented by first defining $G(\chi) =D(\chi) W(\chi)$ 
and $G_f(\chi) = f(\chi) D(\chi) W(\chi)$ for each radial window function $W(\chi)$, and then using the following effective weight in the usual Limber approximation (with the 3D power spectrum evaluated at $k = (\ell+1/2)/\chi$):
\begin{multline*}
G_{\ell}^{\mathrm{Limber}} (\chi)= 
G(\chi) + \frac{2\ell^2 + 2\ell -1}{(2\ell-1)(2\ell+3)} G_f(\chi)
- \frac{(\ell-1)\ell}{(2\ell-1)\sqrt{(2\ell-3)(2\ell+1)}} G_f \lb\frac{2\ell-3}{2\ell+1}\chi \rb \\
- \frac{(\ell+1)(\ell+2)}{(2\ell+3)\sqrt{(2\ell+1)(2\ell+5)}} G_f \lb\frac{2\ell+5}{2\ell+1}\chi \rb \, .
\end{multline*}
}
We have implemented this model in order to compare and contrast the performance of our results. 

\begin{figure}[t!]
\centering
\begin{subfigure}[b]{0.48\textwidth}
\includegraphics[width=\linewidth]{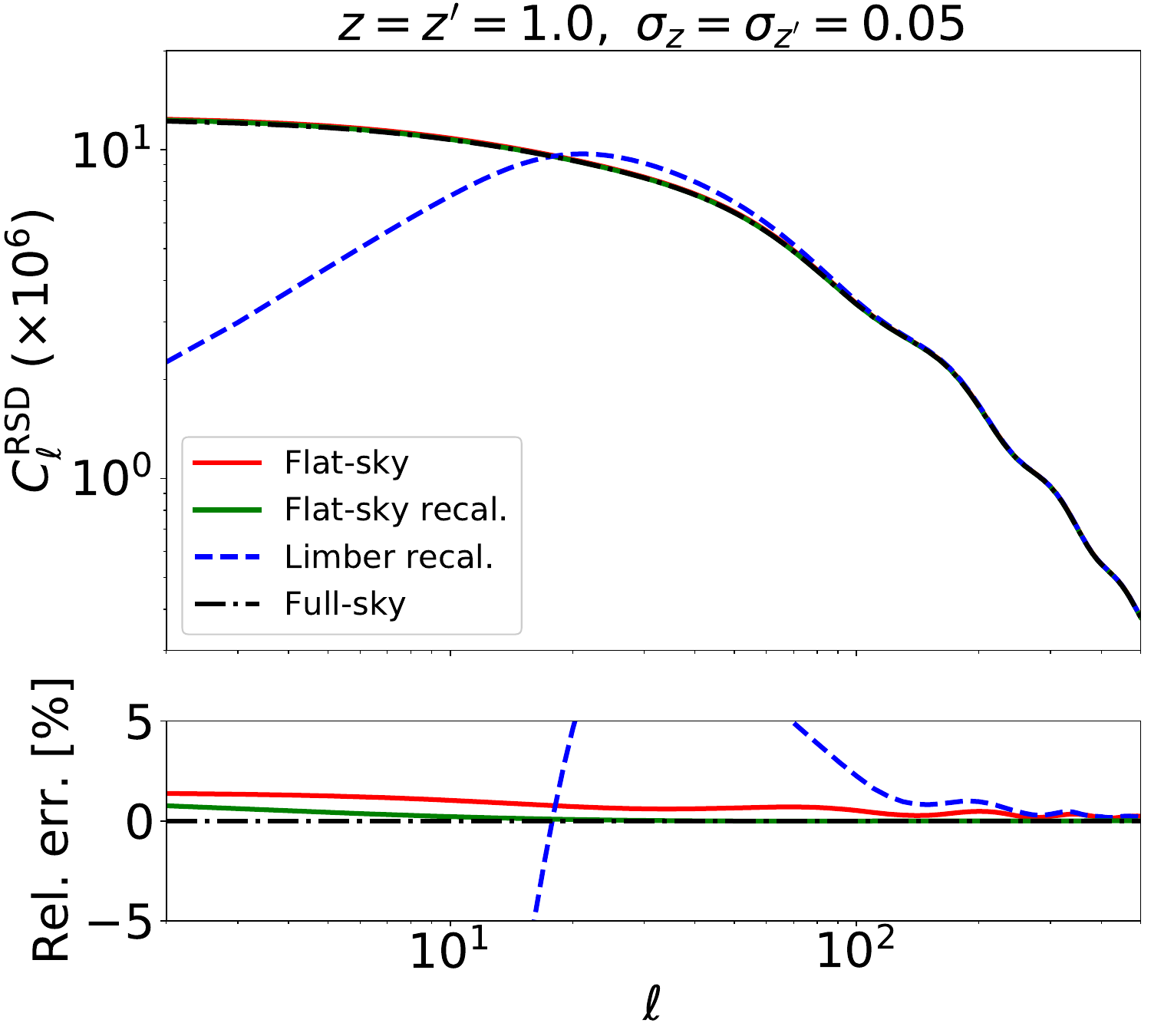}
\label{Cl11_RSD}
\end{subfigure}
\begin{subfigure}[b]{0.48\textwidth}
\includegraphics[width=\linewidth]{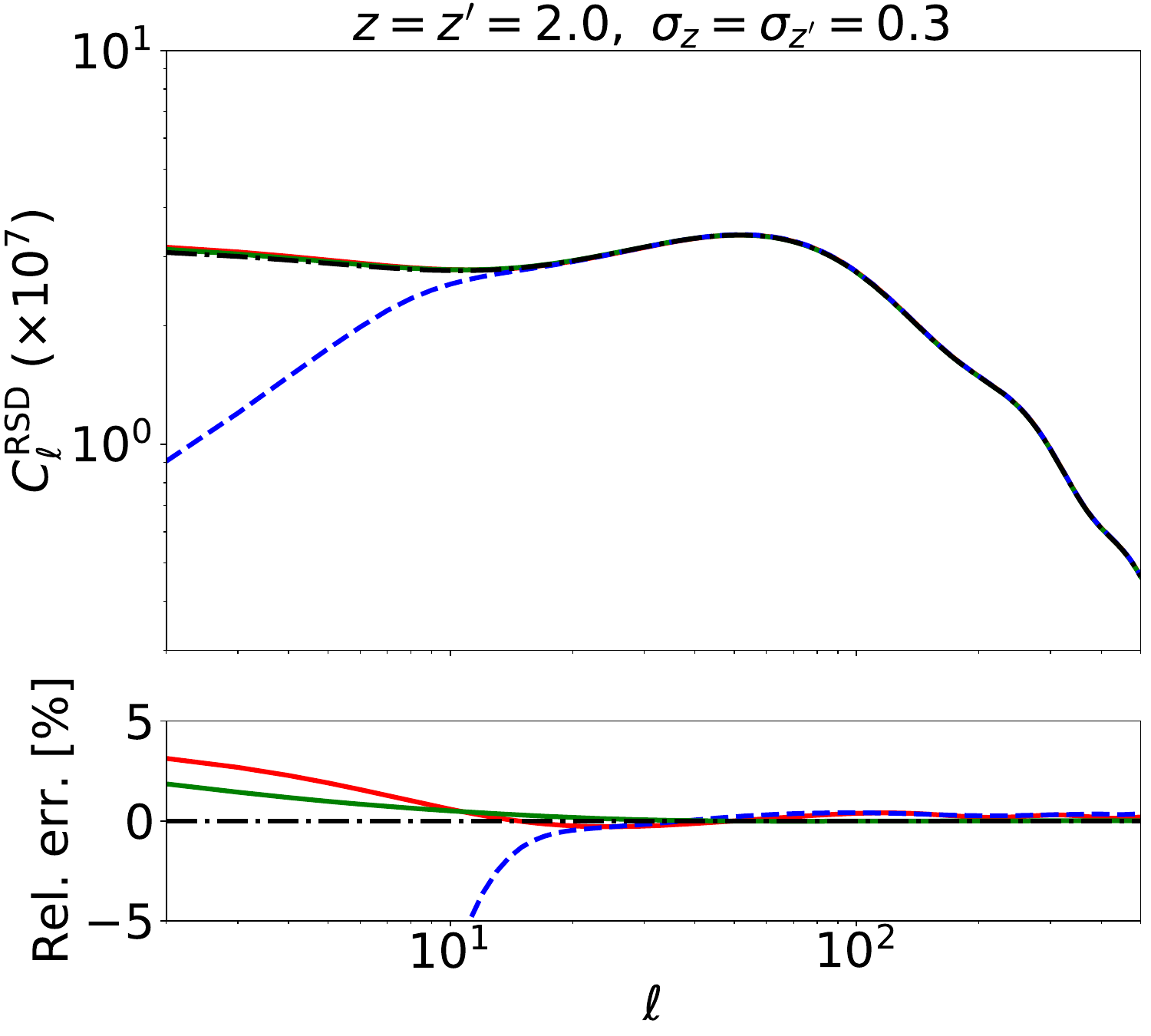}
\label{Cl22_RSD}
\end{subfigure}
\caption{Angular power spectra of projected galaxy clustering, including RSD, with equal-time window functions. 
The left panel is for central redshifts $z=z'=1.0$ and width $\sigma_z = 0.05$, 
while the right panel is for $z=z'=2.0$, with $\sigma_z = 0.3$. 
In both panels, we compare the 
full-sky calculation (black dash-dotted line), the extended Limber approximation (blue dashed line), 
and the flat-sky approximation without geometric recalibration (red solid line) and with recalibration (green solid line).
}
\label{fig:Cell_galaxy_clustering_RSD_auto}
\end{figure}

\begin{figure}[t!]
\centering
\begin{subfigure}[b]{0.48\textwidth}
\includegraphics[width=\linewidth]{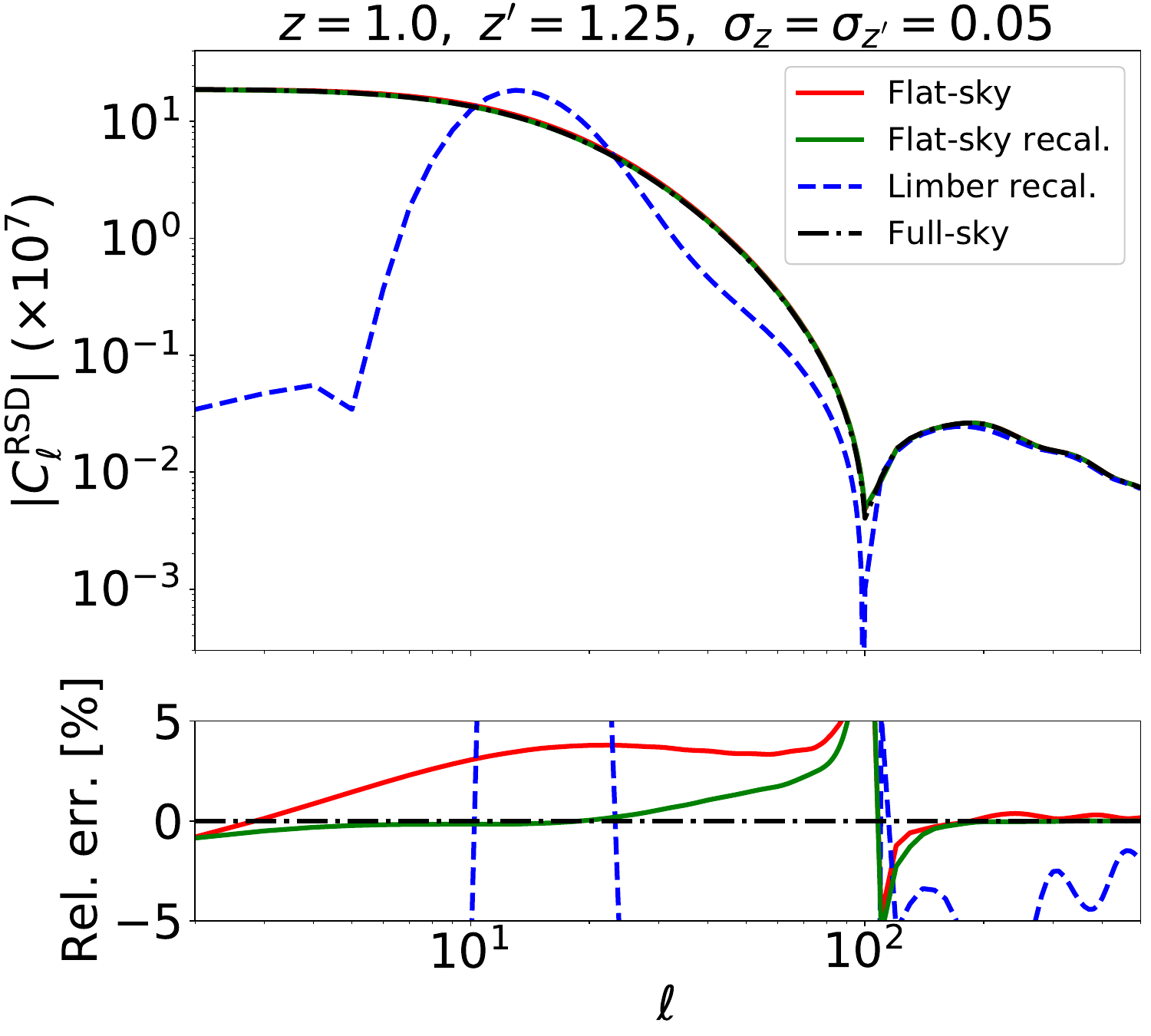}
\end{subfigure}
\hspace{0.1cm}
\begin{subfigure}[b]{0.48\textwidth}
\includegraphics[width=\linewidth]{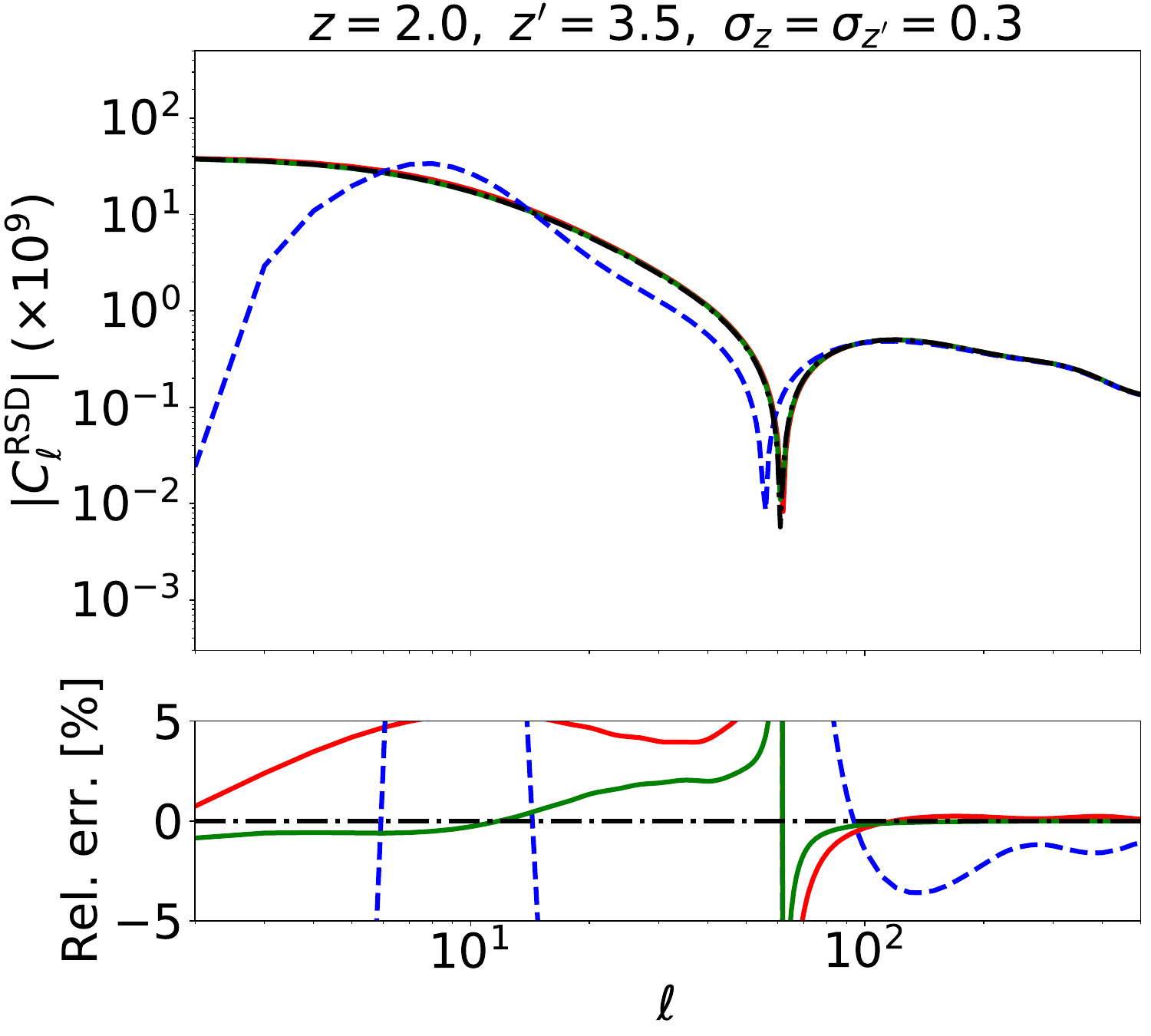}
\end{subfigure}
\caption{As figure~\ref{fig:Cell_galaxy_clustering_RSD_auto} but for unequal-time window functions. 
The left panel is for central redshifts $z=1.0,\ z'=1.25$, with $\sigma_z = 0.05$, while the right panel is for
 $z=2.0,\ z'=3.5$, with $\sigma_z = 0.3$.}
\label{fig:Cell_galaxy_clustering_RSD_cross}
\end{figure}

Figures~\ref{fig:Cell_galaxy_clustering_RSD_auto} and \ref{fig:Cell_galaxy_clustering_RSD_cross} 
show the angular power spectra for projected galaxy clustering, including RSD, for the same equal- and unequal-time setups as in section~\ref{Sec:GalaxyClustering}.  
Our flat-sky approach is in very good agreement with the full-sky results. 
Similar to the case without RSDs, before geometric recalibration, the flat-sky approximation deviates from the full-sky result by less than 
$4\%$ (for the equal-time case) and less than $6\%$ (for the unequal-time case) on all scales of interest. 
Our results again correctly capture the large-scale anti-correlation feature in the unequal-time case.
When the $\ell \to \sqrt{\ell(\ell+1)}$ geometric recalibration is taken into account, our flat-sky results further improve, 
reducing the relative error to less than $2\%$ for both equal-time and unequal-time cases, even on the largest scales.

In comparison, the extended Limber approximation of ref.~\cite{Tanidis:2019} exhibits similar behaviour as without RSD for the case of equal-time window functions (figure~\ref{fig:Cell_galaxy_clustering_RSD_auto}). As expected, the approximation is only accurate at smaller scales and performs better with broader window functions, although the RSD are suppressed for such window functions.
In the case of unequal-time window functions (figure~\ref{fig:Cell_galaxy_clustering_RSD_cross}), with RSD included, the extended Limber approximation does capture the large-scale anti-correlation feature in the angular power spectrum, although the shape is not accurately reproduced.
We note that while the usual Limber approximation without RSD is strictly positive on all scales, the RSD contributions to $G_{\ell}^{\mathrm{Limber}} (\chi)$ of the extended approximation may be negative on large scales for some range of $\chi$, allowing the power to be negative too.
Since RSD make a significant contribution on large scales in figure~\ref{fig:Cell_galaxy_clustering_RSD_cross} (compare with figure~\ref{fig:Cell_galaxy_clustering_cross}), this may account for the rough agreement there between the extended Limber approximation and the full-sky spectrum.
%

\subsection{CMB lensing}
\label{Sec:CMBlensing}

Moving on from the galaxy clustering case, in this subsection
we apply the flat-sky approximation to CMB lensing (see ref.~\cite{Lewis:2006} for a review).
The variable of interest is the CMB lensing potential $\phi$, which is related to the line-of-sight integral of the Newtonian potential $\Psi$ (or, more carefully, the Weyl potential) in the Born approximation:
\begin{equation}
\phi = -2 \int_{0}^{\chi_\ast} d\chi \, \frac{\chi_\ast-\chi}{\chi_\ast \chi} \Psi(\vx; \eta_0-\chi) \, ,
\end{equation}
where $\chi_*$ is the comoving distance of the CMB last-scattering surface
(note that in all the calculations, we assume a flat universe).

Two differences emerge, therefore, when comparing the galaxy clustering and CMB lensing calculations, as follows. 
\begin{enumerate}
\item[(i)] The choice of window function:
\eeq{
\label{Eq:window_lensing}
W_{\Psi}(\chi) = -2\frac{\chi_*-\chi}{\chi_*\chi}\, .
}
\item[(ii)] The field being considered is the Newtonian potential $\Psi$ rather than the matter over-density. They are related by the Poisson equation (in Fourier space and assuming anisotropic stress is negligible)
\eeq{
-k^2\Psi(\vk; z) = 4\pi G a^2\bar{\rho}_m(z) \delta(\vk; z)\, ,
}
where $\bar{\rho}_m(z)$ is the background matter density.
\end{enumerate}
It is straightforward to relate the 3D power spectra of $\Psi$ and $\delta$ according to
\eeq{
P_\Psi(k;\, \chi, \chi') = \frac{3\Omega_m(\chi)a^2(\chi) H^2(\chi)}{2}
\frac{3\Omega_m(\chi')a^2(\chi') H^2(\chi')}{2}\frac{P_\delta(k;\, \chi, \chi')}{k^4}\, ,
}
where $\Omega_m(\chi)$ is the fraction of the energy density in matter, and $H(\chi)$ is the Hubble parameter.
Defining $N_{\Psi}(\chi) \equiv 3\Omega_m(\chi)a^2(\chi) H^2(\chi)/2$, our flat-sky approximation to the angular power spectrum of the lensing potential is
\eq{
C^{\phi\phi}(\ell) = \int d\bar{\chi} d \delta \chi \;  
& W_\Psi\lb \bar{\chi} + \tfrac{1}{2} \delta \chi \rb W_\Psi\lb \bar{\chi} - \tfrac{1}{2} \delta \chi \rb 
N_{\Psi}\lb \bar{\chi} + \tfrac{1}{2} \delta \chi \rb N_{\Psi}\lb \bar{\chi} - \tfrac{1}{2} \delta \chi \rb \non\\
& \times \frac{1}{\bar{\chi}^2} \int \frac{d k_{\parallel}}{2\pi} ~ e^{ i k_{\parallel} \delta \chi}
 \frac{1}{k^4} P_\delta \left(k;\, \bar{\chi}, \delta \chi \right) \, ,
 \label{Eq:phipower}
}
where the total wavenumber $k$ is again given by eq.~\eqref{Eq:lk}. 

For the 3D power spectrum of the matter over-density, $P_\delta$, we again take the unequal-time linear theory 
prediction given in eq.~\eqref{Eq:P_linear}.
After performing the FFTLog expansion of $p(k)$, all we need is to shift every $\nu_i$ in 
eqs.~(\ref{Eq:Expansion}) and~(\ref{Eq:SpecialFunction}) to $\nu_i-4$, while keeping the $\alpha'_i$ unchanged. 
The angular power spectrum of the lensing potential can then be written as
\eq{
C^{\phi\phi}(\ell) = \int d\bar{\chi} d \delta \chi \;  
& W_\Psi\lb \bar{\chi} + \tfrac{1}{2} \delta \chi \rb W_\Psi\lb \bar{\chi} - \tfrac{1}{2} \delta \chi \rb 
\mathbb{C}^{\text{flat}}(\ell,\bar{\chi},\delta \chi) \, ,
\label{eq:Cllensing}
}
where the appropriate unequal-time angular power spectrum for lensing is
\begin{align}
\mathbb{C}^{\text{flat}}(\ell, \chi, \delta\chi) &= 
N_{\Psi}\lb \bar{\chi} + \tfrac{1}{2} \delta \chi \rb N_{\Psi}\lb \bar{\chi} - \tfrac{1}{2} \delta \chi \rb 
\frac{1}{\bar{\chi}^2} \int \frac{d k_{\parallel}}{2\pi} ~ e^{ i k_{\parallel} \delta \chi}
 \frac{1}{k^4} P_\delta \left(k;\, \bar{\chi}, \delta \chi \right) \nonumber \\
&= \frac{1}{\bar \chi^2} N_{\Psi}\lb \bar{\chi} + \tfrac{1}{2} \delta \chi \rb N_{\Psi}\lb \bar{\chi} - \tfrac{1}{2} \delta \chi \rb 
\sum_i \alpha'_i \, \tilde{\ell}^{\nu_i-3} M^{(2)}_{\nu_i-4}\big( \tilde{\ell}\delta\chi \big) \, .
\label{eq:curlyCllensing}
\end{align}

In the CMB lensing literature, when employing the Limber approximation, the angular power spectrum of the lensing convergence $\kappa$ 
is generally calculated rather than the lensing potential $\phi$. The convergence is related to the angular Laplacian of the lensing potential: $\kappa = - \nabla_\theta^2 \phi /2$, so that the spherical power spectra are related by $C_\ell^{\kappa \kappa} = [\ell(\ell+1)]^2C_\ell^{\phi\phi}/4$.
In the flat-sky approximation, we have
\eeq{
\nabla_{\theta}^2\phi = \int d\chi \, W_\Psi (\chi) \chi^2 \nabla_{\bot}^2 \Psi \, , \qquad
\text{where} \qquad
\nabla_{\bot}^2\Psi = \nabla^2 \Psi-\frac{\partial^2 \Psi}{\partial\chi^2} \, .
\label{eq:flatkappa}
}
The field being projected is, therefore, the transverse Laplacian of $\Psi$, which in Fourier space is
\begin{equation}
-k_{\bot}^2\Psi = -\left( k^2-k_{\parallel}^2\right)\Psi 
= -\left( 1- k_{\parallel}^2/k^2\right)k^2\Psi = \left( 1-k_{\parallel}^2/k^2\right)N_{\Psi}(\chi)\delta(k)\, .
\end{equation}
If we calculate the angular power spectrum of $\kappa$ directly with our flat-sky approximation, we have
\begin{multline}
C^{\kappa\kappa}(\ell) = \frac{1}{4}\int d\bar{\chi} d \delta \chi \,
W_\Psi\lb \bar{\chi} + \tfrac{1}{2} \delta \chi \rb W_\Psi\lb \bar{\chi} - \tfrac{1}{2} \delta \chi \rb 
N_{\Psi}\lb \bar{\chi} + \tfrac{1}{2} \delta \chi \rb  N_{\Psi}\lb \bar{\chi} - \tfrac{1}{2} \delta \chi \rb
\bar{\chi}^4 \left(1-\delta^2\right)^2 \\
\times \frac{\tilde{\ell}^4}{\bar{\chi}^2} \int \frac{d k_{\parallel}}{2\pi} \, e^{ i k_{\parallel} \delta \chi} 
\left. \frac{P_\delta(k;\, \bar{\chi},\delta\chi)}{k^4}\right|_{k = \sqrt{k_\parallel^2+\tilde{\ell}^2}} \, ,
\label{Eq:kappapower}
\end{multline}
where $\tilde{\ell} = \sqrt{\ell (\ell+1)}/\chi_g$ with $\chi_g = \bar{\chi}\sqrt{1-\delta^2} = \sqrt{\chi\chi'}$ as usual (after geometric recalibration). Here, the factor $\bar{\chi}^4 \left(1-\delta^2\right)^2 = (\chi\chi')^2$ comes from the conversions between angular and transverse Laplacians at radii $\chi$ and $\chi'$, and $\tilde{\ell}^4$ from $k_\bot^4$. Their product is $[\ell(\ell+1)]^2$ and so, comparing with eq.~\eqref{Eq:phipower}, we recover
\begin{equation}
C^{\kappa\kappa}(\ell) = [\ell(\ell+1)]^2 C^{\phi\phi}(\ell) / 4\, ,
\end{equation}
in agreement with the full-sky relation.

\begin{figure}
    \centering
    \includegraphics[width = 0.96\textwidth]{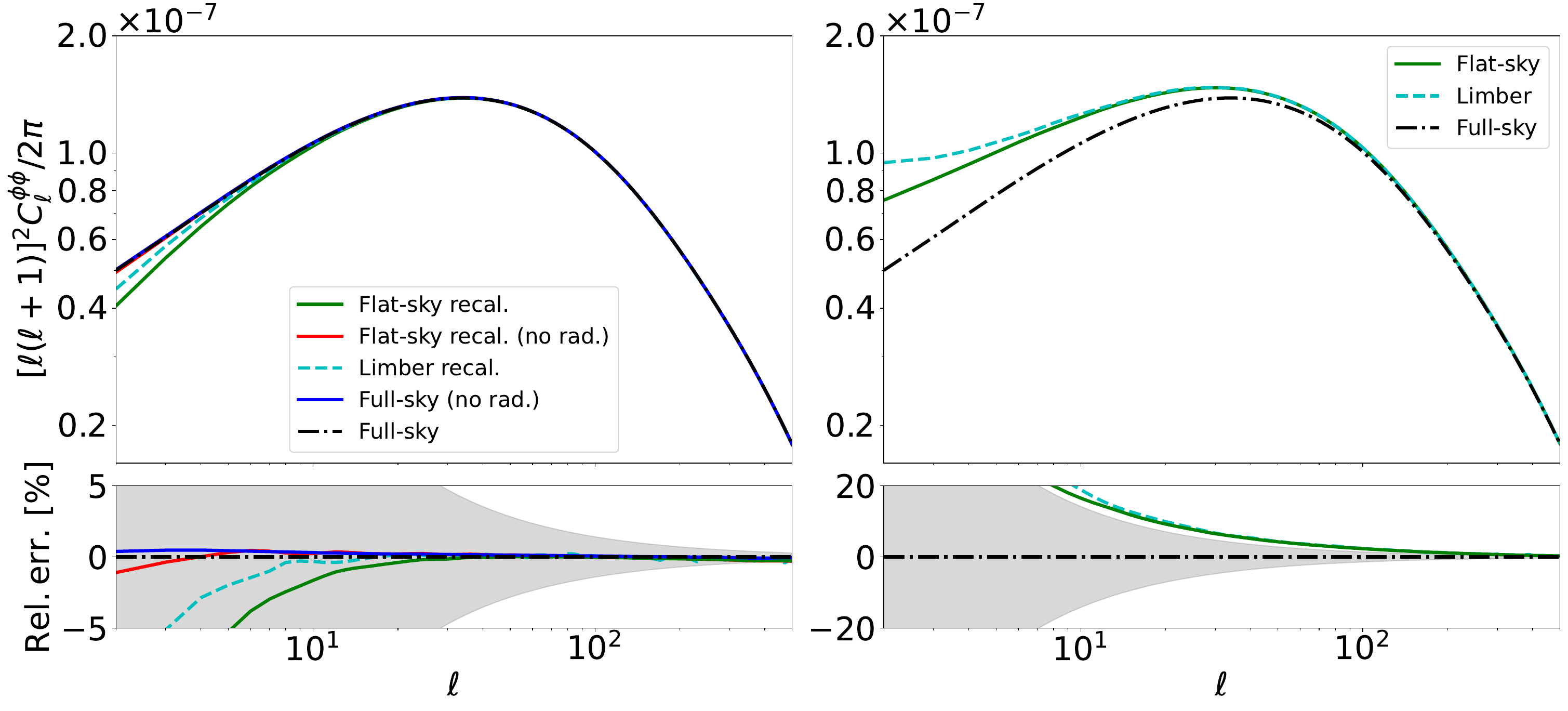}
    \caption{Angular power spectrum of the CMB lensing potential. 
    In the left panel, 
    the green solid line is our flat-sky approximation computing the power spectrum of the lensing potential directly (eq.~\ref{Eq:phipower}), while the red solid line is our alternative flat-sky approximation, computing the convergence directly with the full 3D Laplacian rather than just the transverse Laplacian (eq.~\ref{Eq:kappapowerradialderivs}). The cyan dashed line is the Limber approximation. This, and the flat-sky results in this panel, include geometric recalibration. The black dot-dashed line is the exact full-sky angular power spectrum, while the blue solid line is from the full-sky convergence power spectrum but computed with the full 3D Laplacian, rather than the transverse Laplacian, acting on the gravitational potential. The latter wrongly includes radial derivatives of the potential, but introduces only a very small error.
    In the right panel, we compare the exact full-sky result to the flat-sky (working from the lensing potential) and Limber approximations without geometric recalibration. Comparing with the same colour lines in the left panel illustrates the importance of recalibration (i.e., replacing $\ell$ with $\sqrt{\ell(\ell+1)}$) on large scales for lensing. \ZG{The grey bands represent the (fractional) statistical error in the amplitude of the power spectrum, using all multipoles less than $\ell$, in the cosmic-variance limit as described in section~\ref{Sec:GalaxyClustering}.}}
    \label{Fig:CMBlensing}
\end{figure}

As we shall discuss shortly, the flat-sky expression~\eqref{Eq:phipower} for the angular power spectrum of the CMB lensing potential is less accurate on large scales than the clustering results presented in the previous section. This is because of the different scale dependencies of the 3D power spectra $P(k)$ (gravitational potential versus over-density), with the lensing case having most power on large scales, and the window functions, with lensing being very broad and extending to $\chi=0$ where $W_\Psi \propto 1/\chi$. We propose an alternative flat-sky approximation 
whereby we ignore the distinction between the transverse and full Laplacian in the lensing convergence, approximating
\begin{equation}
\kappa \approx  -\frac{1}{2} \int_{0}^{\chi_\ast} d\chi \, W_\Psi(\chi) \chi^2 \nabla^2 \Psi(\vx; \eta_0-\chi) \, .
\label{eq:kapparadialderivs}
\end{equation}
In this case, the scale dependence of the field being projected is the same as for clustering. The approximation wrongly includes radial derivatives of the gravitational potential. We test the accuracy of this approximation firstly in the full-sky case. In figure~\ref{Fig:CMBlensing} (left panel), we compare the full-sky angular power spectrum of the approximate convergence (eq.~\ref{eq:kapparadialderivs}) with the full-sky $[\ell(\ell+1)]^2 C_\ell^{\phi\phi}/4$; the maximum difference is below $0.5\%$ (and much less on smaller scales).
The effect of the radial derivatives is very small, even on large scales, because of the integration over the broad lensing window function.
We provide further insight into this approximation in appendix~\ref{app:kapparadialderivs}.

We therefore proceed to compute the angular power spectrum of the approximate convergence in eq.~\eqref{eq:kapparadialderivs} using the flat-sky approximation. This is simply given by eq.~\eqref{Eq:kappapower}, but with $1/k^4$ in the integral over $k_\parallel$ replaced with $1/\tilde{\ell}^4$, so that 
\begin{multline}
C^{\kappa\kappa}(\ell) = \frac{1}{4}\int d\bar{\chi} d \delta \chi \,
W_\Psi\lb \bar{\chi} + \tfrac{1}{2} \delta \chi \rb W_\Psi\lb \bar{\chi} - \tfrac{1}{2} \delta \chi \rb 
N_{\Psi}\lb \bar{\chi} + \tfrac{1}{2} \delta \chi \rb  N_{\Psi}\lb \bar{\chi} - \tfrac{1}{2} \delta \chi \rb 
\bar{\chi}^4 \left(1-\delta^2\right)^2 \\
\times \frac{1}{\bar{\chi}^2} \int \frac{d k_{\parallel}}{2\pi} \, e^{ i k_{\parallel} \delta \chi} 
P_\delta\left(\sqrt{k_\parallel^2+\tilde{\ell}^2};\, \bar{\chi},\delta\chi\right) \, .
\label{Eq:kappapowerradialderivs}
\end{multline}

We shall also compare to the Limber approximation. For CMB lensing this is
\eq{
\label{Eq:CMBlensing_Limber}
C_\ell^{\phi\phi, \mathrm{Limber}} = \int d\bar{\chi} ~ W_\Psi^2\lb \bar{\chi}\rb N^2_{\Psi}\lb \bar{\chi}\rb
 \frac{\bar{\chi}^2}{\ell^4} P_{\delta} (\ell/\bar{\chi};\,\bar{\chi}) \, ,
}
while with geometric recalibration we replace $\ell$ with $\sqrt{\ell(\ell+1)}$.

Figure~\ref{Fig:CMBlensing} compares several approximations to the CMB lensing angular power spectrum with the full-sky result. The flat-sky result in eq.~\eqref{Eq:phipower}, which starts from the lensing potential $\phi$, and the Limber approximation in eq.~\eqref{Eq:CMBlensing_Limber} are accurate at the percent level or better for $\ell > 10$ \emph{after geometric recalibration}. The importance of recalibration is shown in the right panel of the figure; without it, the fractional errors in the flat-sky and Limber approximations are significantly larger at low multipoles and not until $\ell \gtrsim 100$ are they at the percent level. For $\ell \lesssim 10$, the recalibrated flat-sky and Limber approximations are less accurate. As noted above, these larger errors arise from the shape of $P_\Psi(k)$ and the form of the window function for CMB lensing. However, we find that if instead we compute the convergence power spectrum approximating the transverse Laplacian with the full 3D Laplacian (eq.~\ref{Eq:kappapowerradialderivs}), the errors on large scales are reduced to below one percent and are much below this on smaller scales. This provides a uniformly accurate approximation, and outperforms the Limber approximation on all scales.

Curiously, the Limber approximation after geometric recalibration is rather more accurate on large scales than the flat-sky approximation that takes the lensing potential as its starting point (compare the cyan and green lines in figure~\ref{Fig:CMBlensing}). This is only the case after geometric recalibration, which appears to overly suppress the CMB lensing angular power spectrum. 

\subsection{Galaxy lensing}
\begin{figure}
    \centering
    \includegraphics[width = 0.96\textwidth]{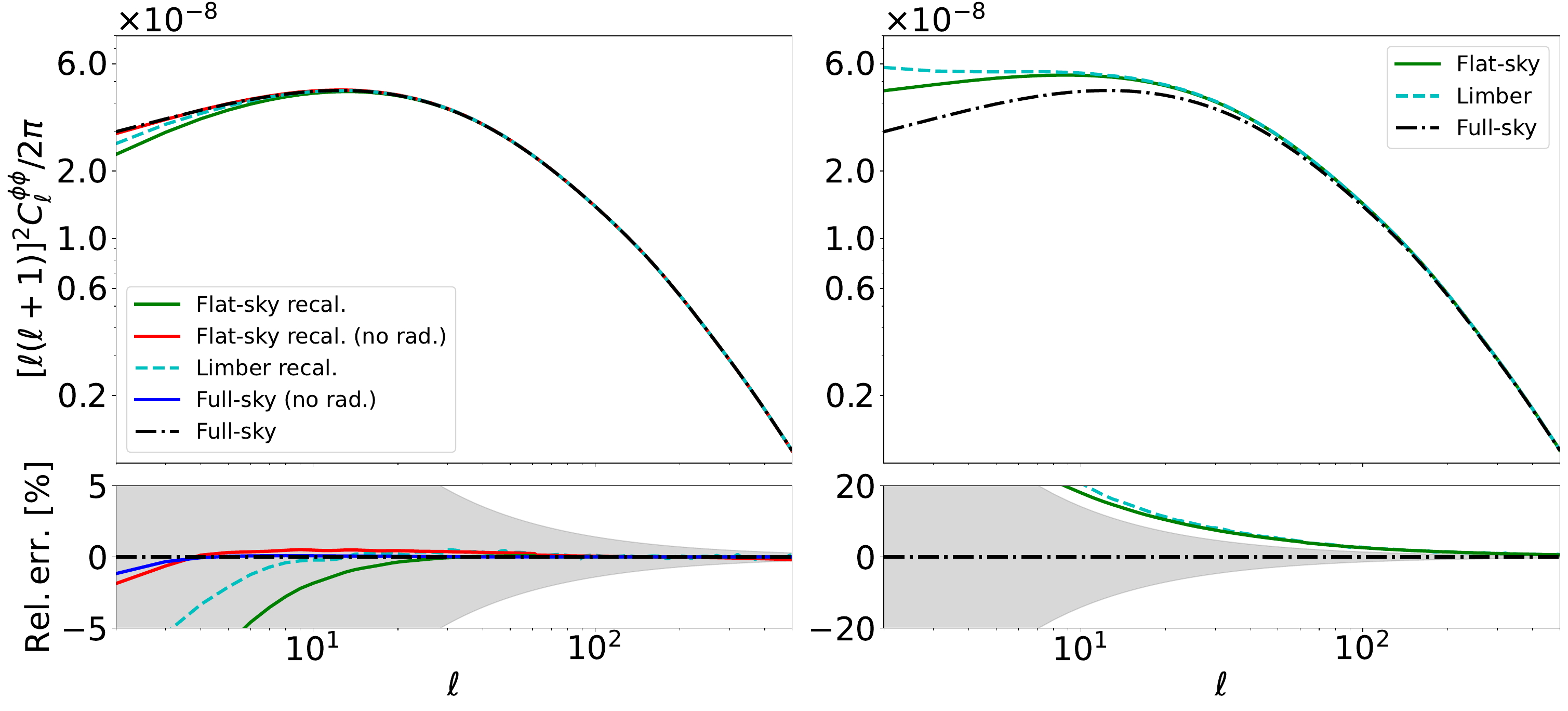}
    \caption{As figure~\ref{Fig:CMBlensing}, but for the lens source plane at $z=2$ rather than the CMB last-scattering surface.}
    \label{Fig:z2_lensing}
\end{figure}

\begin{figure}
    \centering
    \includegraphics[width = 0.6\textwidth]{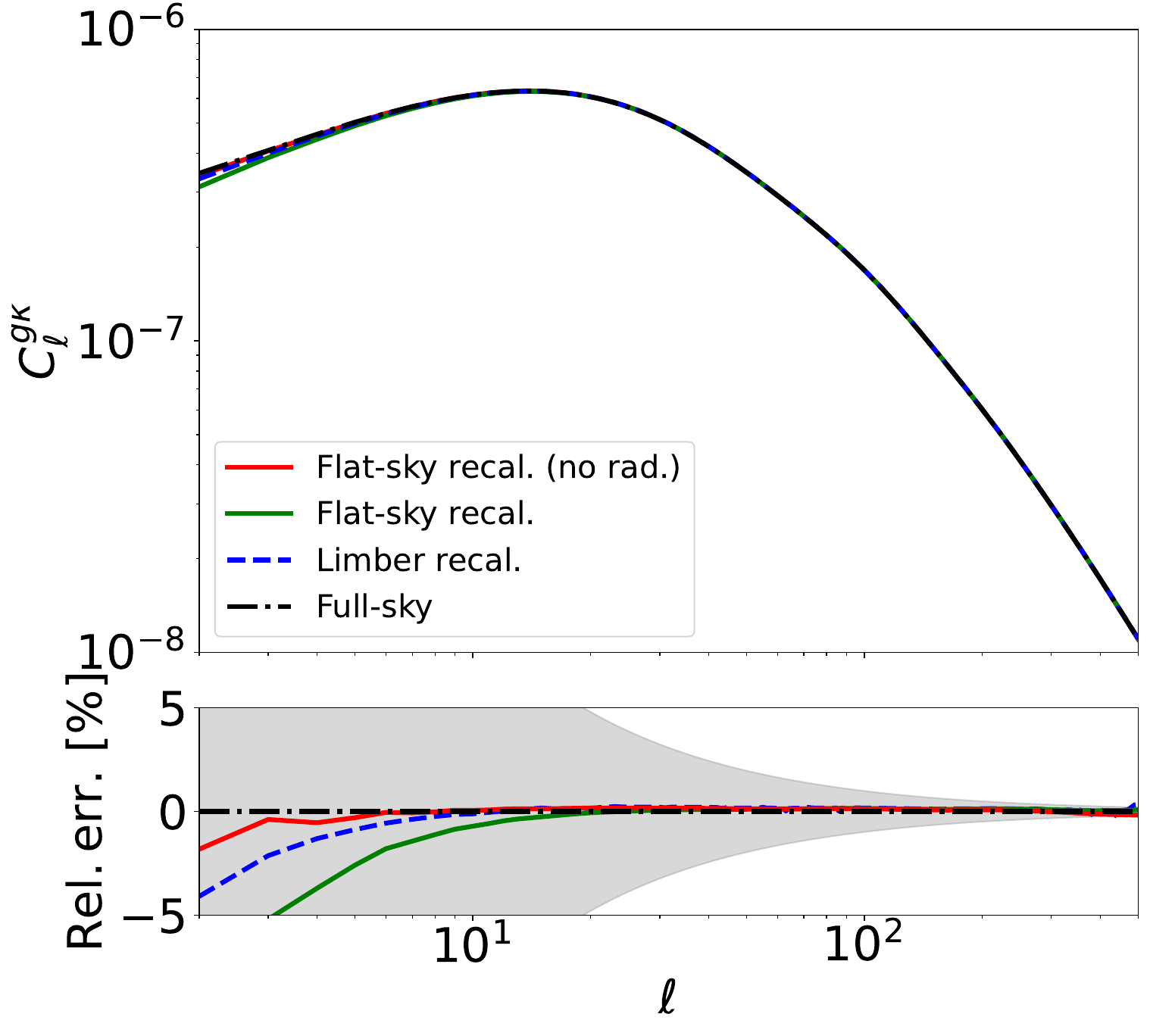}
    \caption{Angular power spectrum for the cross-correlation between the lensing convergence for a source plane at $z=2$ and projected clustering with central redshift $z=0.5$ and width $\sigma_z=0.2$. \ZG{The grey band is derived from the cosmic variance on the amplitude of the cross power spectrum; see eq.~\eqref{eq:crosserrorbound}.}}
    \label{Fig:z2_gk}
\end{figure}

We find similar results as for CMB lensing when the source plane is at lower redshift, as appropriate for lensing of galaxies. In figure~\ref{Fig:z2_lensing}, we show the angular power spectrum of the lensing potential for sources at $z=2$ (compared to $z\approx 1080$ for the CMB last-scattering surface). In figure~\ref{Fig:z2_gk}, we consider the cross-correlation between the lensing convergence field for these sources and projected clustering centred at $z=0.5$ with $\sigma_z=0.2$. In both cases, replacing the transverse Laplacian by the full 3D Laplacian in the spherical convergence is still a very good approximation on all scales. Furthermore, the (recalibrated) flat-sky approximation with the 3D Laplacian very accurately reproduces its full-sky equivalent.

\section{Unequal-time angular power spectra, $\mathbb{C}_\ell$}
\label{Sec:discussion}
In the previous section, we compared the full-sky and flat-sky-approximated projected angular power spectra for
galaxy clustering and CMB lensing. In both cases, the results are given as integrals of the 
unequal-time angular power spectrum $\mathbb C_\ell$ over the observable- and survey-specific window function. 
The expression for the full-sky case is given in eq.~\eqref{Eq:full_sky}, while for the corresponding
flat-sky case we refer to eq.~\eqref{Eq:Cln} (or eq.~\ref{Eq:Cl_geometry}). Any errors in the flat-sky approximation for the projected angular power spectra must therefore\footnote{Of course, errors in the unequal-time angular power spectrum may cancel when performing the radial integrals. Reproducing the unequal-time spectrum for all radii is therefore sufficient, but not necessary, for the observable spectrum to be accurate. For example, with broad window functions the Limber approximation is accurate on small scales, but the unequal-time angular power spectrum will generally still have support for $\delta \chi \neq 0$.} arise from errors in the unequal-time spectra $\mathbb C_\ell$. Moreover, $\mathbb C_\ell$ is entirely independent of the specific survey and functional forms of the chosen window functions and thus solely determined by cosmology. In this section, we analyse these unequal-time spectra in more detail, comparing 
the full-sky $\mathbb{C}_{\ell}$, given in eq.~\eqref{Eq:curly_full}, and the 
flat-sky $\mathbb{C}^{\text{flat}}(\ell)$, given in eq.~\eqref{Eq:curly_flat}. For CMB lensing, we consider the calculation of the lensing potential, wherein we project the gravitational potential, to highlight the effect of the very different scale dependence of the 3D power spectrum. The relevant flat-sky $\mathbb{C}^{\text{flat}}(\ell)$ is given in eq.~\eqref{eq:curlyCllensing}.
Throughout this section, we adopt the geometric recalibration of the 
flat-sky spectra, replacing $\ell$ with $\sqrt{\ell(\ell+1)}$. 

\begin{figure}[t!]
\centering
\includegraphics[width=\linewidth]{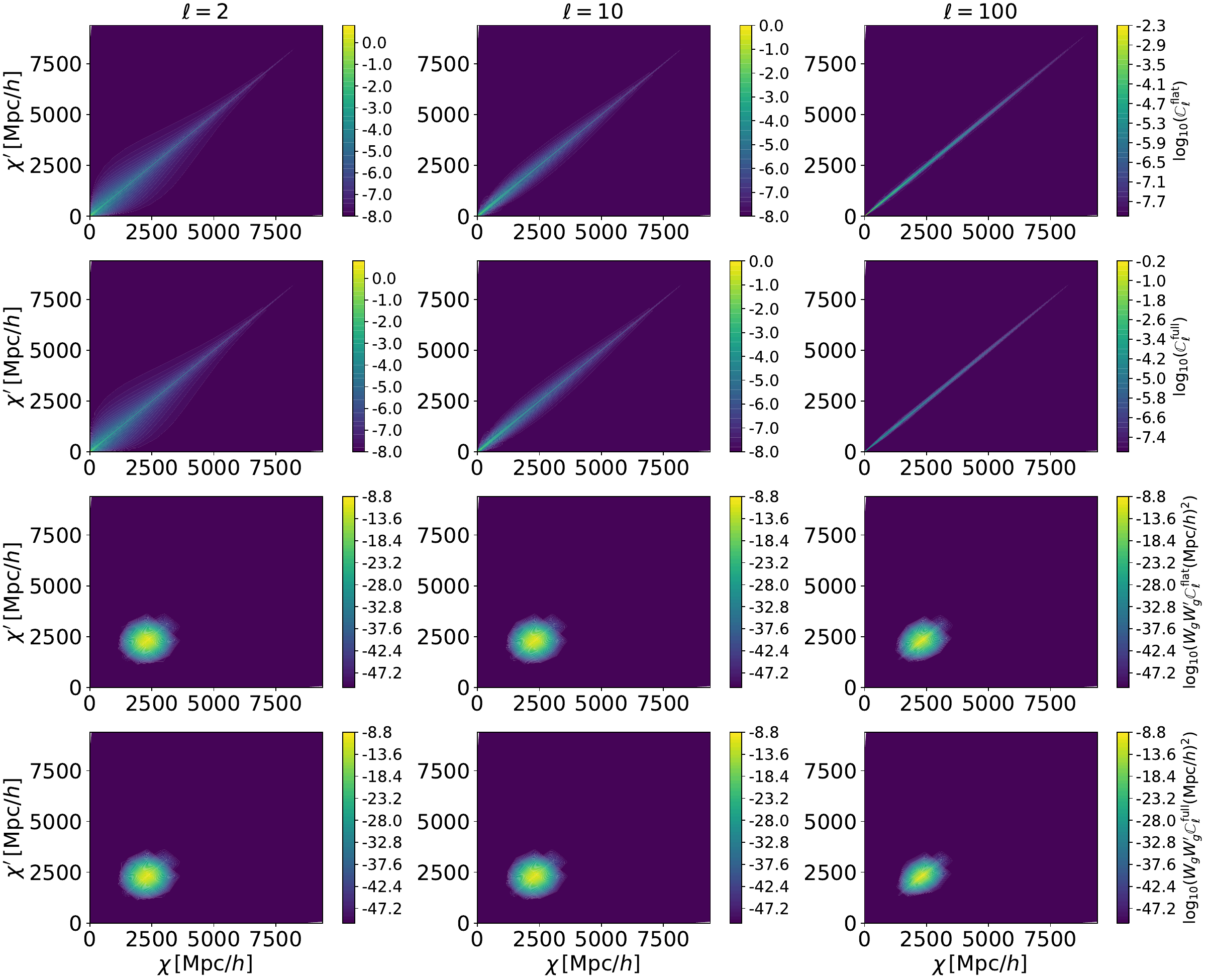}
\caption{2D plots of the unequal-time angular power spectrum $\mathbb{C}_\ell(\chi,\chi')$ for projected clustering.
The columns correspond to multipoles $\ell=2$, $10$ and $100$.
The first row shows the flat-sky approximation $\mathbb{C}^{\mathrm{flat}}(\ell)$, 
while the second row is the full-sky $\mathbb{C}_\ell^{\mathrm{full}}$. 
The third and fourth rows are the flat-sky and full-sky $\mathbb{C}_\ell(\chi,\chi')$, respectively, multiplied by 
radial window functions that are Gaussians centred at $z=z'=1.0$, with widths $\sigma_z=\sigma_{z'}=0.05$. Note the colour scales are logarithmic and the dynamic range is greatly expanded in the bottom two rows to probe the tails of the window functions.
}
\label{Fig:curly_3D_clustering}
\end{figure}

Recall that the key differences between the calculations for matter clustering and the CMB lensing potential are:
(i) the scale dependence of the 3D power spectra $P(k)$ (over-density versus gravitational potential), with the lensing case having most power on large scales; and (ii) the window functions, with lensing being very broad and extending to $\chi=0$ where $W_\Psi \propto 1/\chi$. Both differences would be expected to exacerbate errors in the flat-sky approximation at low multipoles, as seen in figure~\ref{Fig:CMBlensing}. Only the difference in the 3D power spectra impacts the $\mathbb{C}_\ell$ directly, but the window functions control what ranges of $\chi$ and $\chi'$ (or $\bar{\chi}$ and $\delta \chi$) contribute significantly to the projected spectra $C_\ell$.

\begin{figure}[t!]
\centering
\includegraphics[width=\linewidth]{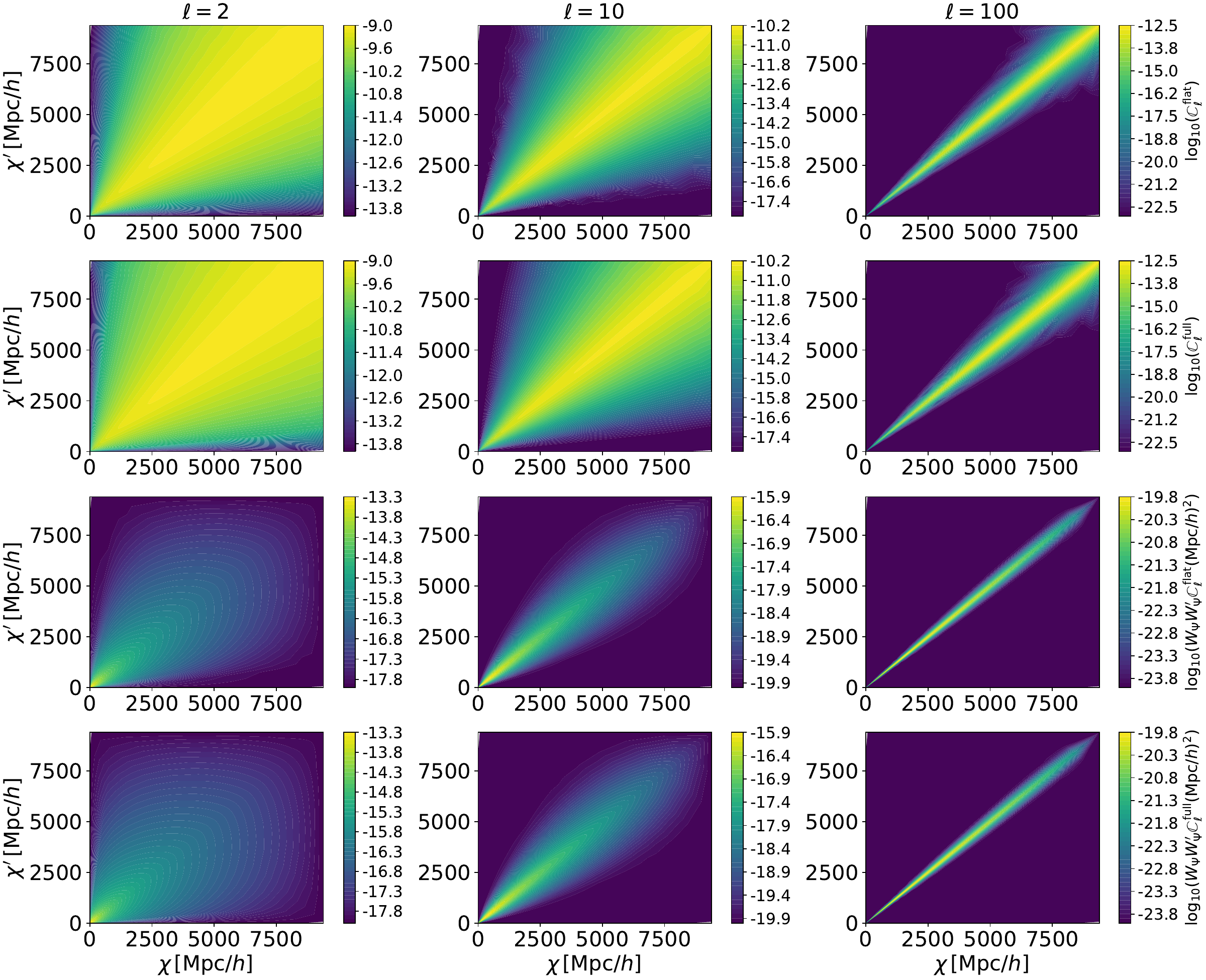}
\caption{As figure~\ref{Fig:curly_3D_clustering} but for the CMB lensing case, where the relevant 3D power spectrum is $P_\delta(k)/k^4$ rather than $P_\delta(k)$. In the bottom two rows, the lensing kernels (eq.~\ref{Eq:window_lensing}) are included.
}
\label{Fig:curly_3D_lensing}
\end{figure}

In figures~\ref{Fig:curly_3D_clustering} (for clustering) and \ref{Fig:curly_3D_lensing} (lensing), we plot in the top two rows
$\mathbb C_\ell^{\rm full}$ and $\mathbb C^{\rm flat}(\ell)$ as a function of $\chi$ and $\chi'$ for multipoles $\ell = 2$, $10$ and $100$.
Comparing the two figures, we see clearly how the dependence of $\mathbb C_\ell$ on distances is 
determined by the shape of the 3D power spectrum. For clustering,
the support of $\mathbb C_\ell$ is strongly clustered around $\chi=\chi'$ ($\delta \chi = 0$) and so correlations between different radii fall sharply with their separation. For window functions broad compared to the off-diagonal width of $\mathbb C_\ell$, the Limber approximation will be very accurate. In constrast, for the lensing case there are significant correlations between widely separated radii, with the distributions becoming broader at low multipoles. These differences arise from the very different $k_\parallel$ dependence of the 3D spectra $P(\sqrt{k_\parallel^2+\tilde{\ell}^2})$ when projecting $\delta$ or $\Psi$, being much more concentrated around $k_\parallel = 0$ for the lensing case. Since the $\delta\chi$ dependence of $\mathbb{C}_\ell$ is determined mostly by the Fourier transform of the 3D spectrum with respect to $k_\parallel$, the narrower 3D spectra for lensing give broader correlations in $\delta \chi$. The bottom two rows in figures~\ref{Fig:curly_3D_clustering} and \ref{Fig:curly_3D_lensing} multiply the $\mathbb{C}_\ell$ by the appropriate pair of window functions; the integral of the resulting quantity over $\chi$ and $\chi'$ gives the observable projected spectra. For clustering, the window functions are Gaussians centred on $z=1$ with widths $\sigma_z = 0.05$. At low multipoles, the lensing angular power spectrum receives significant contributions from nearby structures, a point we shall return to below.
For both clustering and lensing, we see good agreement between the full- and flat-sky results, and 
indeed at the level of figures~\ref{Fig:curly_3D_clustering} and \ref{Fig:curly_3D_lensing}
it is difficult to see any differences.

\begin{figure}[t!]
\centering
\includegraphics[width=\linewidth]{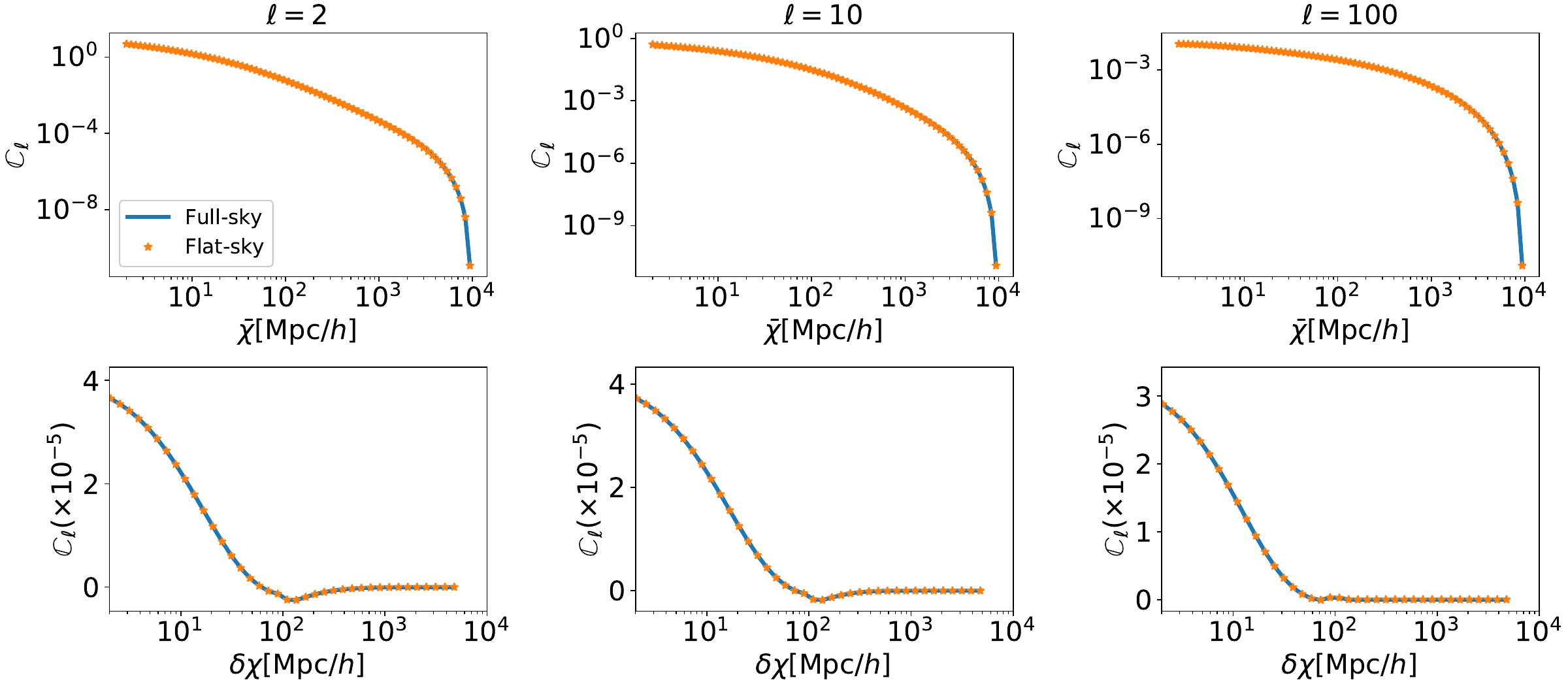}
\caption{Slices through the unequal-time angular power spectrum $\mathbb{C}_\ell(\chi,\chi')$ for projected clustering.
The columns correspond to multipoles $\ell=2$, $10$ and $100$.
The first row compares the flat-sky approximation $\mathbb{C}^{\mathrm{flat}}(\ell)$ and the full-sky $\mathbb{C}_\ell^{\mathrm{full}}$
as a function of $\bar{\chi} \equiv (\chi+\chi')/2$ at $\delta \chi \equiv \chi - \chi' = 0$.
The second row shows slices at $\bar{\chi}=2381\, \mathrm{Mpc}/h$ (around $z=1$) as a function of $\delta\chi$.}
\label{Fig:curly_1D_clustering}
\end{figure}

\begin{figure}[t!]
\centering
\includegraphics[width=\linewidth]{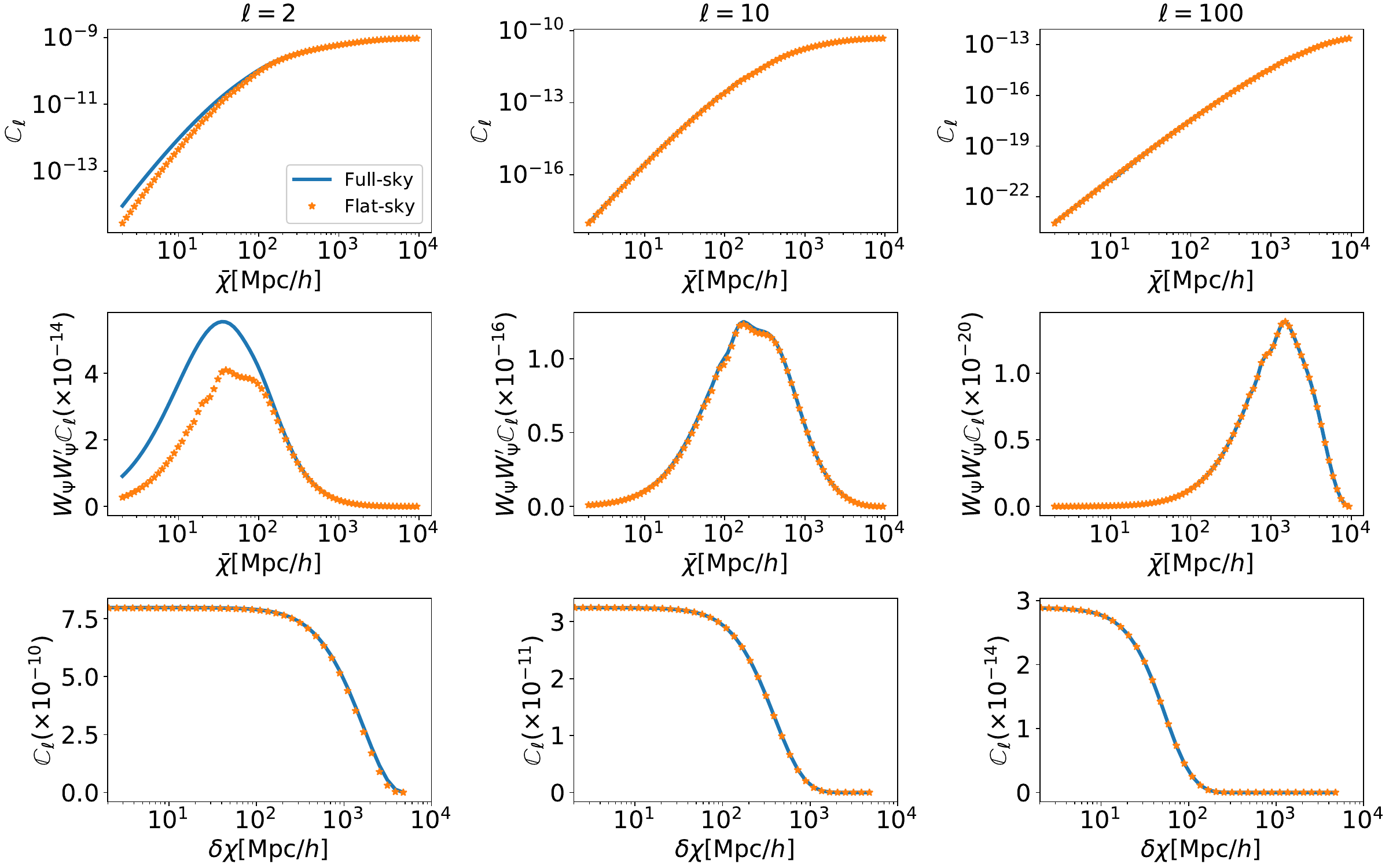}
\caption{Slices through the unequal-time angular power spectrum $\mathbb{C}_\ell(\chi,\chi')$ appropriate to the CMB lensing potential.
The columns correspond to multipoles $\ell=2$, $10$ and $100$.
The first row compares the flat-sky approximation $\mathbb{C}^{\mathrm{flat}}(\ell)$ and the full-sky $\mathbb{C}_\ell^{\mathrm{full}}$
as a function of $\bar{\chi}$ at $\delta \chi = 0$.
The second row multiplies these by the lensing window functions [so the quantity plotted has dimensions of $(h/\mathrm{Mpc})^2$].
The third row shows slices (without multiplication by the window functions) at $\bar{\chi}=2381\, \mathrm{Mpc}/h$ (around $z=1$) as a function of $\delta\chi$.}
\label{Fig:curly_1D_lensing}
\end{figure}

In order to highlight specific aspects of the the unequal-time angular power spectra, we plot
slices through $\mathbb{C}_\ell(\chi,\chi')$ in figures~\ref{Fig:curly_1D_clustering} (clustering) and \ref{Fig:curly_1D_lensing} (lensing).
The top rows show the dependence on $\bar{\chi} \equiv (\chi+\chi')/2$ at $\delta \chi \equiv \chi-\chi'=0$.
For the clustering case, $\mathbb{C}_\ell$ monotonically decreases with $\bar{\chi}$,
while for lensing it monotonically increases.
In section~\ref{subsec:asymptotics} we explain how the asymptotic behaviours of the $\mathbb{C}_\ell$ at small and large $\bar{\chi}$, for $\delta \chi = 0$, follow from the shape of the specific 3D power spectrum being projected.
The bottom rows of the figures show the changes in the compactness of $\mathbb{C}_\ell$ considered as a
function of $\delta \chi$ for a fixed $\bar{\chi}=2381\,\mathrm{Mpc}/h$ (corresponding to $z=1.05$). 
In both clustering and lensing cases, the peak of $\mathbb{C}_\ell$ over $\delta\chi$ becomes narrower as $\ell$ increases so that correlations between different radii are suppressed. As also seen in the 2D plots in figures~\ref{Fig:curly_3D_clustering} and \ref{Fig:curly_3D_lensing}, the peak is much broader and flat-topped for lensing than clustering.
It is also noteworthy that the clustering $\mathbb{C}_\ell$ can be negative for low multipoles, giving anti-correlations between different radii, but the lensing $\mathbb{C}_\ell$ cannot. This behaviour arises from the 3D clustering power spectrum, $P_\delta(\sqrt{k_\parallel^2+\tilde{\ell}^2})$, having a local minimum at $k_\parallel = 0$ for $\tilde{\ell} \ll k_{\text{eq}}$, where $k_{\text{eq}}$ is the matter-radiation equality scale.
%

Let us focus next on the comparison of the full-sky and flat-sky $\mathbb C_\ell$ in the sectional plots in figures~\ref{Fig:curly_1D_clustering} (clustering) and~\ref{Fig:curly_1D_lensing} (CMB lensing). At low multipoles, differences start to appear in the lensing case for small radii $\chi$ and $\chi'$ or, equivalently, at low $\bar{\chi}$ for $\delta\chi=0$ (as in the top row of figure~\ref{Fig:curly_1D_lensing}), or for $|\delta \chi| \approx 2 \bar{\chi}$ (bottom row of the figure). These differences are exacerbated by multiplication by the lensing radial window functions, which, recall, behave as $W_\Psi \propto 1/\chi$ for $\chi \ll \chi_\ast$, as shown in the second row of the figure. At the lowest multipoles, significant errors in $\mathbb{C}^{\text{flat}}(\ell)$ from lenses at $\bar{\chi} \lesssim 200\,\text{Mpc}/h$ lead to the relatively large errors in the angular power spectrum of the CMB lensing potential seen in figure~\ref{Fig:CMBlensing}. However, we emphasise again that a simple and accurate work-around is to calculate the flat-sky convergence power spectrum, replacing the transverse Laplacian by the full 3D Laplacian.

\begin{figure}
\centering
\includegraphics[width=\linewidth]{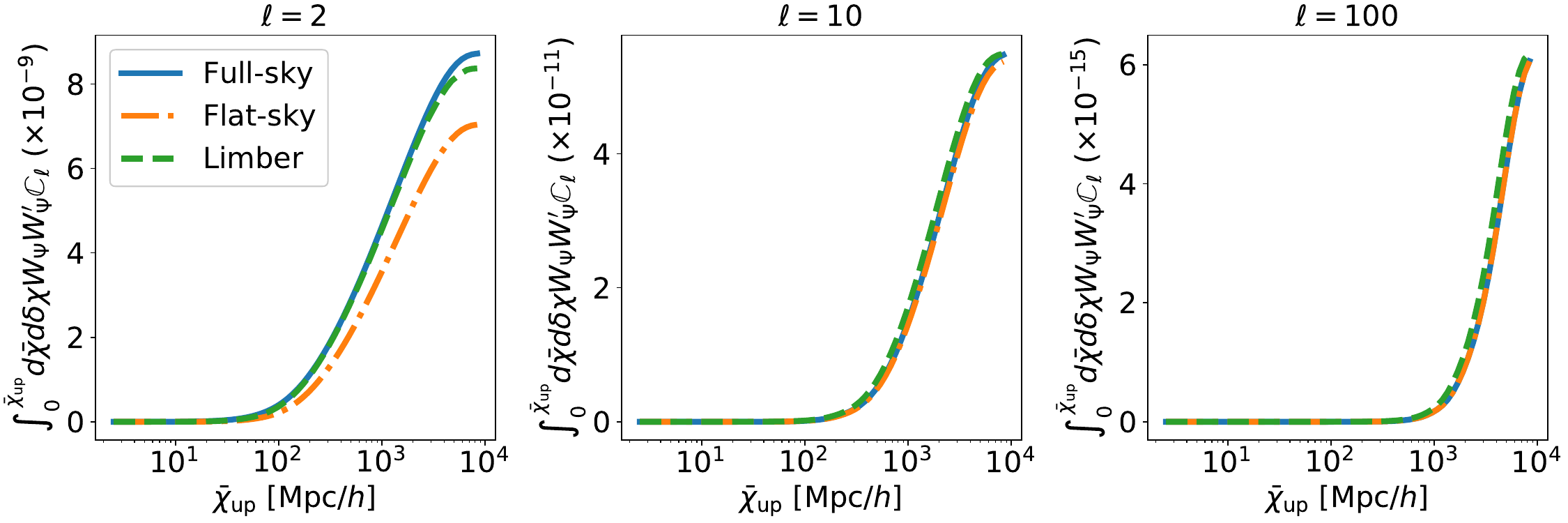}
\caption{Cumulative contribution to the angular power spectrum of the CMB lensing potential, $C^{\phi\phi}_\ell$, from $\bar{\chi} \leq \bar{\chi}_{\text{up}}$. Results are shown for the full-sky (blue solid lines), Limber approximation (green dashed lines) and flat-sky approximation (eq.~\ref{eq:Cllensing}; orange dot-dashed lines). Each panel represents a different multipole $\ell$.}
\label{Fig:curly_int_lensing}
\end{figure}

Finally, let us look in more detail at the range of distances that contribute to the CMB lensing angular power spectrum. We show in figure~\ref{Fig:curly_int_lensing} the cumulative contribution to $C^{\phi\phi}_\ell$ from $\bar{\chi} \leq \bar{\chi}_{\text{up}}$ (integrating over all allowed $\delta \chi$). We compare the full-sky result, the Limber approximation and the flat-sky approximation (proceeding via the lensing potential). As expected from the projection, CMB lensing picks up more contributions from larger distances at higher multipoles. At $\ell=2$, around $10\,\%$ of the angular power comes from $\bar{\chi} < 200 \,\text{Mpc}/h$, where figure~\ref{Fig:curly_1D_lensing} shows there are significant errors in the flat-sky approximation.

\subsection{Asymptotic analysis of the flat-sky $\mathbb{C}_\ell$}
\label{subsec:asymptotics}

In this subsection, we discuss the asymptotic behaviour of the unequal-time angular power spectrum for small and large
$\bar \chi$ at $\delta\chi =0$, corresponding to the top rows in figures~\ref{Fig:curly_1D_clustering} and~\ref{Fig:curly_1D_lensing}.
In the earlier sections,
we have found good agreement between the full-sky and the flat-sky behaviour; for simplicity, we therefore focus on analyzing the flat-sky case only.
For this purpose, we also reduce the 3D power spectrum to a very simple form, 
roughly capturing the behaviour of the $\Lambda$CDM matter power spectrum in the IR/UV regime. 
We thus have
\begin{equation}
    k_\text{eq}^3 p(k) \sim
\begin{cases}
    k/k_\text{eq}\, , & k\ll k_{\rm eq}\, , \\
    (k/k_{\text{eq}})^{-3+\epsilon}\, , & k\gg k_{\rm eq} \, ,
\end{cases}
\end{equation}
where $k_{\rm eq}$ is the power spectrum equality scale and $0<\epsilon<1$.

Setting $\delta \chi =0$ in eq.~(\ref{Eq:curly_flat}), $\mathbb{C}^{\text{flat}}(\ell)$ simply 
reduces to integrating the matter power spectrum over $k_\parallel$. We consider first the behaviour as $\bar{\chi} \rightarrow 0$ so that $\tilde{\ell} \gg k_{\text{eq}}$ for all $\ell$. For $k_\parallel \ll \tilde{\ell}$, we approximately consider the power spectrum as independent of $k_\parallel$,
i.e., $k_\text{eq}^3 p(k)\sim (\tilde{\ell}/k_{\text{eq}})^{-3+\epsilon}$, while for $k_\parallel \gg \tilde{\ell}$, 
we can ignore the contribution of $\tilde{\ell}$ in $k$, 
so that $k_\text{eq}^3 p(k) \sim (k_\parallel/k_{\text{eq}})^{-3+\epsilon}$.
For clustering, therefore, we have 
\begin{align}
\label{Eq:flat_small_chi}
\mathbb {C}^{\mathrm{flat}}({\ell}) 
&\sim \frac{D^2(\bar{\chi})}{k_\text{eq}^3\bar{\chi}^2}\left[ \int_0^{\tilde{\ell}} \frac{d k_{\parallel}}{2\pi} \, \left(\frac{\tilde{\ell}}{k_{\text{eq}}}\right)^{-3+\epsilon}
+ \int_{\tilde{\ell}}^{\infty} \frac{d k_{\parallel}}{2\pi} \, \left(\frac{k_\parallel}{k_{\text{eq}}}\right)^{-3+\epsilon} \right] \non \\
&= \frac{1}{2\pi} \left(\frac{3-\epsilon}{2-\epsilon}\right) [\ell(\ell+1)]^{-1+\epsilon/2}(k_\text{eq}\bar{\chi})^{-\epsilon}\, ,
\end{align}
where we have used the asymptotic property of the linear growth factor at low $\bar{\chi}$, $D(\bar{\chi})\to \text{const.}$
The only change in the lensing case is that we divide $P(k)$ by an additional factor of $k^4$, 
which gives the asymptotic $\bar{\chi}^{4-\epsilon}$ behaviour. 
These asymptotic behaviours can be observed in the top rows of figures~\ref{Fig:curly_1D_clustering} and~\ref{Fig:curly_1D_lensing} for projected clustering and lensing, respectively.

Similarly, for large $\bar{\chi}$, we can take $\tilde{\ell}$ to be much smaller than $k_{\mathrm{eq}}$. 
We have to consider three regions in the matter power spectrum. For $k_\parallel \ll \tilde{\ell}$, 
we still have a constant $k_\text{eq}^3 p(k) \sim \tilde{\ell}/k_\text{eq}$. For $\tilde{\ell} \ll k_\parallel \ll k_{\rm eq}$, 
we take $k_\text{eq}^3 p(k)\sim k_\parallel/k_\text{eq}$, while for $k_\parallel \gg k_{\rm eq}$, 
we take $k_\text{eq}^3 p(k) \sim (k_\parallel/k_\text{eq})^{-3+\epsilon}$. Collecting the three regimes gives us
\begin{align}
\label{Eq:flat_large_chi}
\mathbb {C}^{\mathrm{flat}}(\ell) 
&\sim \frac{D^2(\bar{\chi})}{k_\text{eq}^3\bar{\chi}^2}\left[ \int_0^{\tilde{\ell}} \frac{d k_{\parallel}}{2\pi} \, \left(\frac{\tilde{\ell}}{k_{\text{eq}}}\right)
+ \int_{\tilde{\ell}}^{k_\text{eq}} \frac{d k_{\parallel}}{2\pi} \, \left(\frac{k_\parallel}{k_{\text{eq}}}\right)
+ \int_{k_\text{eq}}^\infty \frac{d k_{\parallel}}{2\pi} \, \left(\frac{k_\parallel}{k_{\text{eq}}}\right)^{-3+\epsilon}
\right] \non \\
&= \frac{1}{2\pi} \frac{D^2(\bar{\chi})}{(k_\text{eq}\bar{\chi})^2}\left( \frac{\ell(\ell+1)}{2(k_\text{eq}\bar{\chi})^2}
 +\frac{1}{2} + \frac{1}{2-\epsilon} \right)\, .
\end{align}
For projected clustering and $k_\text{eq} \bar{\chi} \gg 1$, this implies that 
$\mathbb {C}^{\mathrm{flat}}_{\ell} \sim D^2(\bar{\chi}) \bar{\chi}^{-2}$,
consistent with the steeper fall-off in the top row of figure~\ref{Fig:curly_1D_clustering} at large $\bar{\chi}$. 
Following the same calculation for lensing (and dividing the power spectrum by $k^4$), 
we have
\begin{align}
\label{Eq:flat_large_chi_lensing}
\mathbb {C}^{\mathrm{flat}}(\ell) 
&\sim \frac{1}{k_\text{eq}^3\bar{\chi}^2}\left[ \int_0^{\tilde{\ell}} \frac{d k_{\parallel}}{2\pi} \, \left(\frac{\tilde{\ell}}{k_{\text{eq}}}\right)^{-3}
+ \int_{\tilde{\ell}}^{k_\text{eq}} \frac{d k_{\parallel}}{2\pi} \, \left(\frac{k_\parallel}{k_{\text{eq}}}\right)^{-3}
+ \int_{k_\text{eq}}^\infty \frac{d k_{\parallel}}{2\pi} \, \left(\frac{k_\parallel}{k_{\text{eq}}}\right)^{-7+\epsilon}
\right] \non \\
&= \frac{1}{2\pi} \frac{1}{(k_\text{eq}\bar{\chi})^2}\left( \frac{3}{2}\frac{(k_\text{eq}\bar{\chi})^2}{\ell(\ell+1)}
 -\frac{1}{2} + \frac{1}{6-\epsilon} \right)\, ,
\end{align}
where we have already used the property that $D(\bar{\chi})N_{\Psi}(\bar{\chi})$ is approximately constant at large $\bar{\chi}$. This means that $\mathbb{C}^\text{flat}(\ell)$ approaches to a constant for large $\bar{\chi}$, 
as we observe in the top row of figure~\ref{Fig:curly_1D_lensing}.

\section{Conclusion}
\label{Sec:conclusion}

We revisited the performance of the flat-sky approximation in evaluating the angular power spectrum of projected fields. Unlike the commonly used Limber approximation, we retained the contribution of wave-modes along the line of sight (labelled by $k_\parallel$). We showed that with this inclusion, very accurate spectra can be obtained with the flat-sky approximation including cases where the projection is with narrow radial window functions or window functions with limited overlap. Moreover, the approximation generally remains accurate at low multipoles, where curved-sky effects are expected to be significant, with a simple geometric recalibration of the flat-sky results. Our flat-sky approximation also provides a self-consistent way of including redshift-space distortion effects, given that we explicitly retain the dependence on the $k_\pp$ modes.

Our results were developed in a general form, independent of the specifics of a given observable or survey. For this purpose, we separated the computation of the projected angular power spectrum into computation of the unequal-time angular power spectrum, $\mathbb C_\ell(\chi, \chi')$, and integration over survey- and observable-specific radial window functions, $W(\chi)$ and $W'(\chi')$. The $\mathbb C_\ell(\chi, \chi')$
includes only the projections of the relevant unequal-time 3D power spectrum onto radii $\chi$ and $\chi'$, and takes the form of a Fourier integral over $k_\parallel$ with conjugate variable $\delta \chi = \chi-\chi'$. Given the specifics of each observable and survey, the treatment of the radial integrals over the window functions can then be further optimised.

A key step in our numerical evaluation of the angular power spectrum in the flat-sky approximation is accelerating the evaluation of $\mathbb C_\ell(\chi, \chi')$ based on a discreet Mellin transform of the 3D power spectrum, commonly called the FFTLog algorithm. We utilise FFTLog to expand the 3D power spectrum as a sum of power-laws in $k$ with complex frequencies $\nu_i$. The Fourier integral required for $\mathbb C_\ell(\chi, \chi')$ can then be expressed as a sum of modified Bessel functions of the second kind $K_{\nu_i}(x)$. In comparison, an equivalent procedure in the calculation of the full-sky angular power spectrum results in a sum of terms involving the hypergeometric functions ${}_2F_1(a,b;c;z)$~\cite{Assassi2017}, increasing the computational complexity and evaluation time. Moreover, these modified Bessel functions depend only on the variable $x = \delta\chi \sqrt{\ell(\ell+1)} /\sqrt{\chi \chi'}$. We can therefore easily pre-compute and store the $K_{\nu_i}(x)$ functions and interpolate them as required to evaluate the angular power spectrum at different points in parameter space when sampling over cosmological parameters during inference. For simplicity, we presented results using only the linear-theory 3D power spectrum. However, the method is general and can be applied to any choice of the 3D power spectrum (not necessarily given in perturbative form).
We note that further numerical optimisations are possible, e.g., implementing an optimised set of the complex frequencies $\nu_i$ and improved sampling for the integration over the radial window functions. We are exploring the former, and emulation of the associated Mellin transform, in ongoing work. Even so, and irrespective of these further optimisations, our flat-sky approach already provides a significant computational improvement compared to the full-sky FFTLog-based algorithm.

We presented results for projected clustering (e.g., galaxy surveys) with and without redshift-space distortions, gravitational lensing and their cross-correlation. In general, we found excellent (percent-level or better) agreement of the flat-sky and full-sky results on all scales (including also the lowest $\ell$s), with the flat-sky results outperforming the Limber approximation. However, we found rather larger errors from the flat-sky approximation when evaluating the angular power spectrum of the CMB lensing potential at low multipoles. These errors arise since the power spectrum of the gravitational potential, whose projection determines the CMB lensing potential, is very red with most power on large scales. Furthermore, the radial window function rises as $1/\chi$ at small radial distances, so at low multipoles non-negligible contributions to the lensing power spectrum come from nearby lenses. We presented a simple work-around that restores the accuracy of the flat-sky approximation on large scales. This involves working with an approximate form of the lensing convergence, which approximates the transverse part of the Laplacian of the gravitational potential (coming from the angular Laplacian in the convergence) by the full 3D Laplacian including radial derivatives. In this manner, the approximate convergence involves the projection of the matter over-density, which has a power spectrum that peaks at the equality scale $k_\text{eq}$. As an aside, we verified that using an equivalent approximation in the spherical convergence is very accurate at all multipoles. \ZG{We found that, in all scenarios the accuracy of recalibrated flat-sky approximation is well with in the range of cosmic variance.}

To conclude, in this paper, we investigated the performance of the flat-sky approximation taking into account the correlations along the line of sight. This gives an accurate and efficient approximation of the full-sky angular power spectrum (at sub-percent level) for galaxy clustering (including redshift-space distortions) and gravitational lensing. We have developed an efficient Python implementation of our flat-sky method, based on the FFTLog expansion, which we make publicly available.\footnote{https://github.com/GZCPhysics/BeyondLimber.git} With this method and further optimisation, one can compute the projected angular power spectrum at speeds comparable to the Limber approximation. This opens an alternative route for efficient computation of angular power-spectrum observables in parameter searches. Such an efficient framework is especially important in light of many upcoming surveys, taking data on large and intermediate scales and at high signal-to-noise ratios. Moreover, effects due to the photon geodesic projections (i.e., relativistic corrections) will be an important consideration for these surveys. The addition of these effects goes beyond our current work; however, incorporating them into our framework should be fairly straightforward. Moreover, and in line with the aforementioned surveys, the cross-correlations of unequal-time narrow window functions are expected to play an increasingly important role in the future cosmological analysis. Thus, our computational framework offers a valuable and efficient means to analyse these upcoming data sets.

\begin{acknowledgments}
We thank William Matthewson, Alvise Raccanelli and Martin White for useful discussions. 
Z.V. acknowledges the support of the Kavli Foundation. A.C.\ acknowledges support from the STFC (grant numbers ST/N000927/1 and ST/S000623/1).
\end{acknowledgments}  

\appendix
\section{Full-sky angular power spectrum in real and redshift space}
\label{app:full_sky_Cell}

This appendix reviews the calculation of the full-sky angular power spectrum using the FFTLog algorithm. 
First, we consider the projected density field without redshift-space distortions (RSD) following ref.~\cite{Assassi2017}. 
The full-sky angular power spectrum is given by eq.~\eqref{Eq:full_sky}, with the unequal-time angular power spectrum $\mathbb{C}_\ell$ given by
eq.~\eqref{Eq:curly_full}. After the FFTLog expansion of the 3D power spectrum, we have
\eeq{
\label{Eq:fullsky_expansion}
\mathbb C_{\ell} = 4\pi DD' \sum_i \alpha_i \int \frac{k^2 dk}{2\pi^2}\; 
k^{\nu_i} j_\ell(k \chi) j_\ell(k \chi')\, ,
}
where, for compactness, we use the notation 
$D = D(\chi)$ and $D' = D(\chi')$ for the linear growth factors (and similarly for $f$, later, when we include redshift-space distortions).
Introducing the variables $v\equiv k\chi$ and $t\equiv \chi'/\chi$, 
the integral over $k$ in eq.~\eqref{Eq:fullsky_expansion} becomes 
\eeq{
\label{Eq:I_nu}
\int \frac{k^2 dk}{2\pi^2}\; 
k^{\nu_i} j_\ell(k \chi) j_\ell(k \chi') = \chi^{-\nu_i-3} \int \frac{dv}{2\pi^2}\; v^{\nu_i+2} j_\ell(v) j_\ell(vt)\, ,
}
which can be evaluated in terms of the hypergeometric function using
\eq{
I_{\ell,\ell'}\lb \nu, t \rb 
&\equiv \int dv\; v^{\nu+2} j_{\ell}(v) j_{\ell'}(vt) \non \\
&= \frac{2^{\nu+2}\pi \Gamma\lb \frac{\ell+\ell'+\nu+3}{2} \rb}{4\Gamma\lb \frac{\ell-\ell'-\nu}{2} \rb \Gamma\lb \ell'+\frac{3}{2} \rb} t^{\ell'}\,_{2}F_{1}\lb \frac{\nu+2-\ell+\ell'}{2}, \frac{\ell+\ell'+\nu+3}{2}; \ell'+\frac{3}{2}; t^2\rb\, ,
\label{Eq:I_l1_l2}
}
assuming $t \leq 1$ (i.e., $\chi' \leq \chi$). 
When encountering $\chi'>\chi$, we instead use the variable transformation 
$v\equiv k\chi'$ and $t\equiv \chi/\chi'$. 
The full-sky angular power spectrum in eq.~\eqref{Eq:full_sky}
is therefore given by\footnote{This expression is symbolic as it does not account for the splitting up of the domain of integration according to whether $\chi' \leq \chi$ or $\chi' > \chi$.}
\eeq{
C_{\ell} = \frac{2}{\pi}  \int d \chi d \chi'\; (DW) (D'W') \lb \sum_i \alpha_i \chi^{-\nu_i-3}
I_{\ell,\ell}\lb \nu_i, t \rb \rb \, ,
\label{eq:full_sky_cell}
}
where $W = W(\chi)$ and $W'=W'(\chi')$.
We use this expression to compute our reference full-sky results presented in section~\ref{Sec:RD}.

If we include RSD, the unequal-time angular power spectrum on the full sky is modified to
\eeq{
\label{Eq:Fullsky_RSD}
\mathbb C_{\ell}\lb \chi, \chi' \rb 
= 4\pi \int \frac{k^2 dk}{2\pi^2}\; P(k;\chi,\chi') 
\lb b\, j_\ell(k \chi)-f j^{(2)}_\ell(k \chi) \rb 
\lb b'\, j_{\ell'}(k \chi')-f' j^{(2)}_{\ell'}(k \chi') \rb,
}
where $j^{(2)}_\ell(k \chi)$ is the second derivative of the spherical Bessel function evaluated at $k \chi$. We have included the galaxy bias, which depends on $k$ and $\chi$ generally, although we have set it to unity in the rest of the paper. Recall, also, that $f \equiv d\ln D / d\ln a$ is the logarithmic growth rate.
To deal with the derivatives of the spherical Bessel functions, ref.~\cite{Assassi2017} made use of integration by parts to reduce all integrals to the form in eq.~\eqref{Eq:I_nu}. However, this moves the derivatives onto the window functions (and growth factors and rates), which may introduce complexities in the case of irregular window functions,
which is often the case for realistic surveys. 

We proceed, instead, by expressing the second derivative of the spherical Bessel function as follows:
\eq{
 j^{(2)}_\ell(k \chi) = 
 - \lb 1 - \frac{\ell (\ell-1) }{k^2 \chi^2} \rb j_\ell(k \chi)
 + \frac{2}{k \chi} j_{\ell+1}(k \chi)\, .
}
Using this expression in eq.~\eqref{Eq:Fullsky_RSD} gives
\begin{multline}
\mathbb C_{\ell} = \int \frac{k^2 dk}{2\pi^2}\; P(k;\chi,\chi') \Bigg[
\lb b+ f - f \frac{\ell (\ell-1)}{k^2 \chi^2} \rb \lb b+f' - f' \frac{\ell (\ell-1) }{k^2 \chi^{'2}} \rb j_\ell(k \chi) j_\ell(k \chi') \\
-  \frac{2f'}{k \chi'}  \lb b+f - f \frac{\ell(\ell-1)}{k^2\chi^2} \rb j_{\ell}(k \chi)j_{\ell+1}(k \chi')
-  \frac{2f}{k \chi} \lb b+f' - f' \frac{\ell(\ell-1)}{k^2\chi'^2} \rb j_{\ell+1}(k \chi)j_{\ell}(k \chi') \\
 + \frac{4ff'}{k^4\chi \chi'} j_{\ell+1}(k \chi)j_{\ell+1}(k \chi') \Bigg]\, ,
\label{Eq:CurlyC_RSD}
\end{multline}
which, after applying the decomposition of $P(k;\chi,\chi')$ as done in eq.~\eqref{Eq:fullsky_expansion},
is fully determined by the integrals given in eq.~\eqref{Eq:I_l1_l2}.

\section{Approximating the CMB lensing convergence}
\label{app:kapparadialderivs}

In section~\ref{Sec:CMBlensing}, we introduced the approximation of using the full 3D Laplacian rather than its transverse part when computing the CMB lensing convergence. Here, we aim to provide further insight into why this is very accurate, even on large scales.

We consider the full-sky case, so that eq.~\eqref{eq:flatkappa} becomes for the true convergence
\eeq{
\kappa = -\frac{1}{2}\int d\chi \, W_\Psi (\chi) \chi^2 \nabla_{\bot}^2 \Psi \, , \qquad
\text{where} \qquad
\nabla_{\bot}^2\Psi = \nabla^2 \Psi- \frac{1}{\chi^2}\frac{\partial}{\partial \chi}\left(\chi^2 \frac{\partial \Psi}{\partial \chi}\right) \, .
}
If instead we replace the transverse (i.e., angular) part of the 3D Laplacian, $\nabla_\bot^2$, with the full Laplacian, $\nabla^2$, we introduce an error, $\Delta \kappa$, involving the radial derivatives:
\begin{equation}
\Delta \kappa = \frac{1}{2} \int_{0}^{\chi_\ast} d\chi \, W_\Psi(\chi) \frac{\partial}{\partial \chi}\left(\chi^2 \frac{\partial \Psi}{\partial \chi}\right)(\vx; \eta_0-\chi) \, .
\end{equation}

It is instructive to consider the case of an Einstein--de Sitter universe, where the potential $\Psi$ does not evolve in time. In that case, the radial derivatives in $\Delta \kappa$ can be replaced by total derivatives along the line of sight. Integrating by parts and using the explicit form for the lensing window function, eq.~\eqref{Eq:window_lensing}, we find
\begin{equation}
\Delta \kappa = - \left[\Psi\right]^{\chi_\ast}_0 \, ,
\end{equation}
i.e., boundary terms evaluated on the last-scattering surface and at the observer. The term at the observer does not contribute for $\ell \geq 1$.
The angular power spectrum of the approximated convergence differs from that of the true convergence by
\begin{equation}
C^{\kappa\kappa,\text{\,approx}}_{\ell}  - C^{\kappa\kappa}_\ell = C_{\ell}^{\Delta\kappa \Delta \kappa}  - 2 C_\ell^{\kappa \Delta \kappa} \, .
\label{eq:kappapowererror}
\end{equation}
The angular power spectrum of $\Delta \kappa$ at low multipoles ($\ell \ll k_{\text{eq}} \chi_\ast$) is
\eeq{
C_{\ell}^{\Delta\kappa \Delta \kappa}  = 4\pi \int \frac{k^2 dk}{2\pi^2}\; P_{\Psi}(k;\, \chi_\ast) \, j^2_\ell(k \chi_\ast) \approx 2\pi \frac{9 A_{\text{s}}}{25 \ell(\ell+1)}\, ,
}
where the approximation assumes a scale-invariant power spectrum of primordial curvature perturbations with amplitude $A_{\text{s}}$. At $\ell=2$, $C_2^{\Delta\kappa\Delta\kappa} \approx 7.5 \times 10^{-10}$ compared to $C_2^{\kappa\kappa} \approx 7.9 \times 10^{-8}$, so the error from the power spectrum of $\Delta \kappa$ is sub-percent. The error term $- 2 C_{\ell}^{\kappa \Delta \kappa} $ from the cross-correlation between $\Delta \kappa$ and $\kappa$ is also expected to be small, even on large scales, since the potential fluctuations on the last-scattering surface are weakly correlated with the fluctuations at lower $\chi$ that dominate the convergence.

\bibliographystyle{JHEP}
\bibliography{main}
\end{document}